%% file: spm4aqp.tex
\documentclass[sigconf,edbt]{acmart-edbt2019}

\usepackage{booktabs}
\usepackage{subcaption}
\usepackage{listings}

\setcopyright{rightsretained}
\acmDOI{}
\acmISBN{XXX-X-XXXXX-XXX-X}
\acmConference[EDBT 2019]{22nd International Conference on Extending Database Technology (EDBT)}{March 26-29, 2019}{Lisbon, Portugal} 
\acmYear{2019}
\renewcommand{\shortauthors}{P. Utama et al.}

\begin{document}

\title{Model-based Approximate Query Processing}

\author{
Moritz Kulessa\textsuperscript{1},
Alejandro Molina\textsuperscript{1}, 
Carsten Binnig\textsuperscript{1,2},
Benjamin Hilprecht\textsuperscript{1},
Kristian Kersting\textsuperscript{1}
\\
\textsuperscript{1} TU Darmstadt, Germany 
\textsuperscript{2} Brown University, USA 
}

\begin{abstract}
Interactive visualizations are arguably the most important tool to explore, understand and convey facts about data. In the past years, the database community has been working on different techniques for Approximate Query Processing (AQP) that aim to deliver an approximate query result given a fixed time bound to support interactive visualizations better. However, classical AQP approaches suffer from various problems that limit the applicability to support the ad-hoc exploration of a new data set: (1)~Classical AQP approaches that perform online sampling can support ad-hoc exploration queries but yield low quality if executed over rare subpopulations. (2)~Classical AQP approaches that rely on offline sampling can use some form of biased sampling to mitigate these problems but require a priori knowledge of the workload, which is often not realistic if users want to explore a new database. 

In this paper, we present a new approach to AQP called Model-based AQP that leverages generative models learned over the complete database to answer SQL queries at interactive speeds. Different from classical AQP approaches, generative models allow us to compute responses to ad-hoc queries and deliver high-quality estimates also over rare subpopulations at the same time. In our experiments with real and synthetic data sets, we show that Model-based AQP can in many scenarios return more accurate results in a shorter runtime. Furthermore, we think that our techniques of using generative models presented in this paper can not only be used for AQP in databases but also has applications for other database problems including Query Optimization as well as Data Cleaning.
\end{abstract}

\maketitle

\input{sections/introduction.tex}
\input{sections/overview.tex}
\input{sections/sum_product_networks.tex}
\input{sections/query_compilation.tex}
\input{sections/probability_based.tex}
\input{sections/sample_based.tex}

\input{sections/experimental_evaluation.tex}

\input{sections/related_work.tex}
\input{sections/conclusion.tex}

\bibliographystyle{abbrv}
\bibliography{bib} 

\input{sections/appendix.tex}

\end{document}

%% file: sections/introduction.tex
\section{Introduction}
\label{sec:intro}

\begin{figure}
\centering
\includegraphics[width=0.47\textwidth]{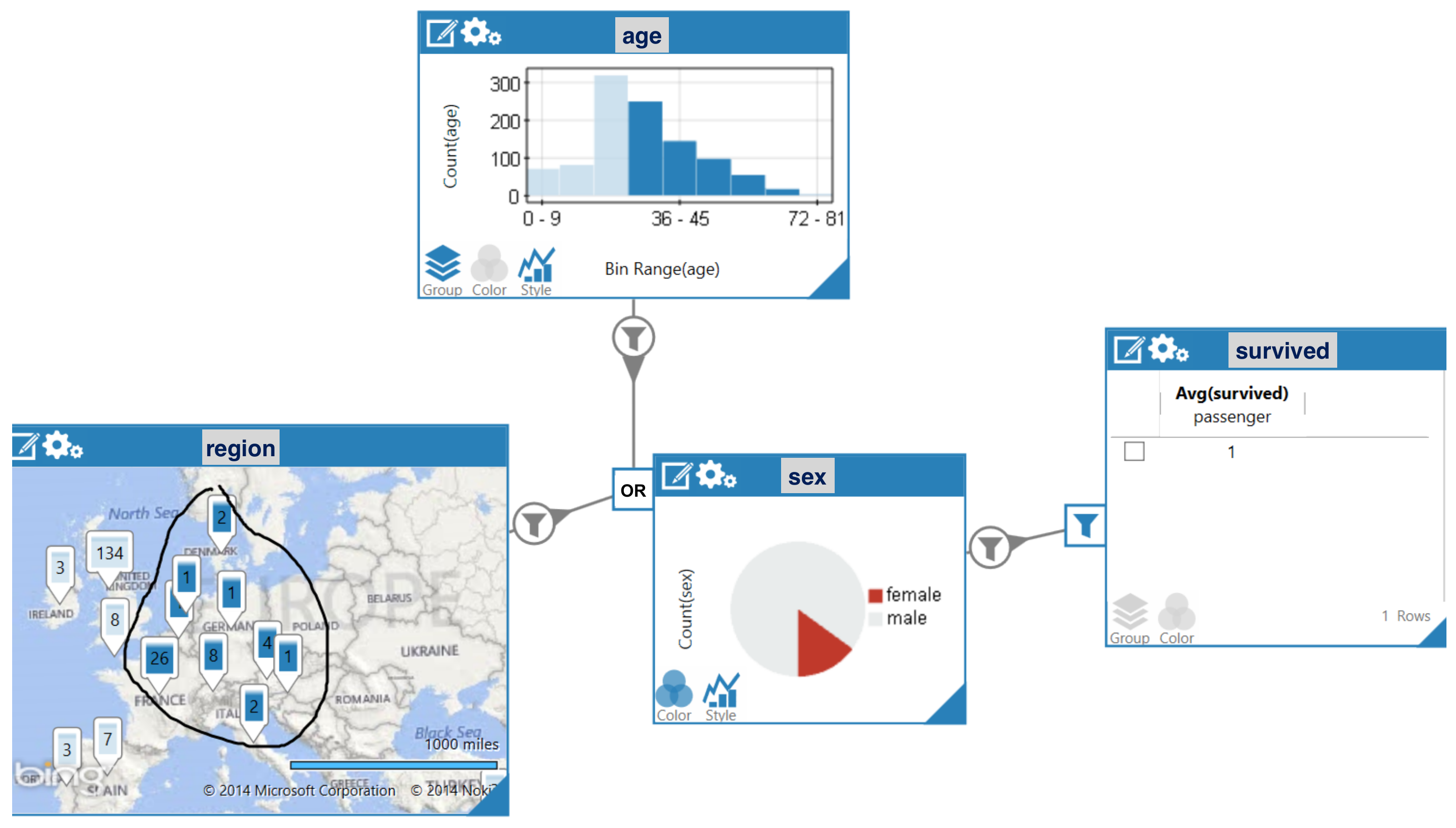}

\caption{Interactive data exploration session.}

\label{fig:panoramic}
\end{figure}

\paragraph{Motivation:} Interactive visualizations are arguably the most important tool to explore, understand and convey facts about data. 
For example, as part of data exploration visualizations are used to quickly skim through the data and look for patterns \cite{Stolte.2002,tableau2012}.
This requires to generate a sequence of visualizations and allow the user to interact with them.
Figure~\ref{fig:panoramic} shows an example screenshot of an interactive data exploration session over the Titanic data set\footnote{\url{http://biostat.mc.vanderbilt.edu/wiki/pub/Main/DataSets/titanic.html}}. 
First, the user analyzes the distribution of passengers by region (leftmost) in a type of a choropleth map visualization, then he looks at the histogram of the age of passengers (topmost).
Afterwards, he selects only those passengers that are from certain countries in Europe or above the age of 25, and looks at the distribution between males and females in form of a pie chart (middle). 
Finally, the user selects only the female passengers and shows the average rate of survived passengers (right). 

All of these scenarios require interactive visualizations that quickly react to a given user interaction (e.g., restricting the data using a given filter condition). 
Unfortunately, when the data sets are larger, computing a single visualization can take seconds or even minutes,  creating a significant barrier to interactive data analysis.
A recent study \cite{DBLP:journals/tvcg/LiuH14} has shown that visual delays of $500$ms tend to decrease both end-user activity and data set coverage, due to the reduction in rates of user interaction that is crucial for overall observation, generalization and hypothesis.
Maybe surprisingly, traditional database systems are ill-suited for speeding up visualizations and can not guarantee interactive response times especially since data sizes are growing constantly. 

The database community has been working on different techniques for Approximate Query Processing (AQP) that aim to deliver an estimate of the query result given a fixed time bound.
In the past, different techniques for AQP have been proposed including approaches that leverage pre-computed samples or synopses as well as techniques that sample from the underlying data at query runtime.
However, all the existing AQP approaches suffer from various limitations that restrict the applicability to support the ad-hoc exploration of a new data set \cite{aqp_reuse}: (1) AQP approaches that are based on online sampling (e.g., DBO \cite{aqp_dobra}, CONTROL \citep{aqp_control}, approXimateDB \cite{aqp_wander_join}) are able support ad-hoc queries on the one hand but on the other hand can only provide good approximations for queries over the mass of the distribution, while queries over rare sub-populations yield results with loose error bounds or even result in missing values in the query results.
(2) AQP approaches that rely on offline sampling can use some form of biased sampling to mitigate this problem (e.g., AQUA \cite{aqp_aqua}, BlinkDB \cite{aqp_blinkdb}), but therefore usually require a priori knowledge of the workload which is often not realistic if users want to explore a new database using ad-hoc queries. 

\begin{figure*}
\centering
\includegraphics[width=\textwidth]{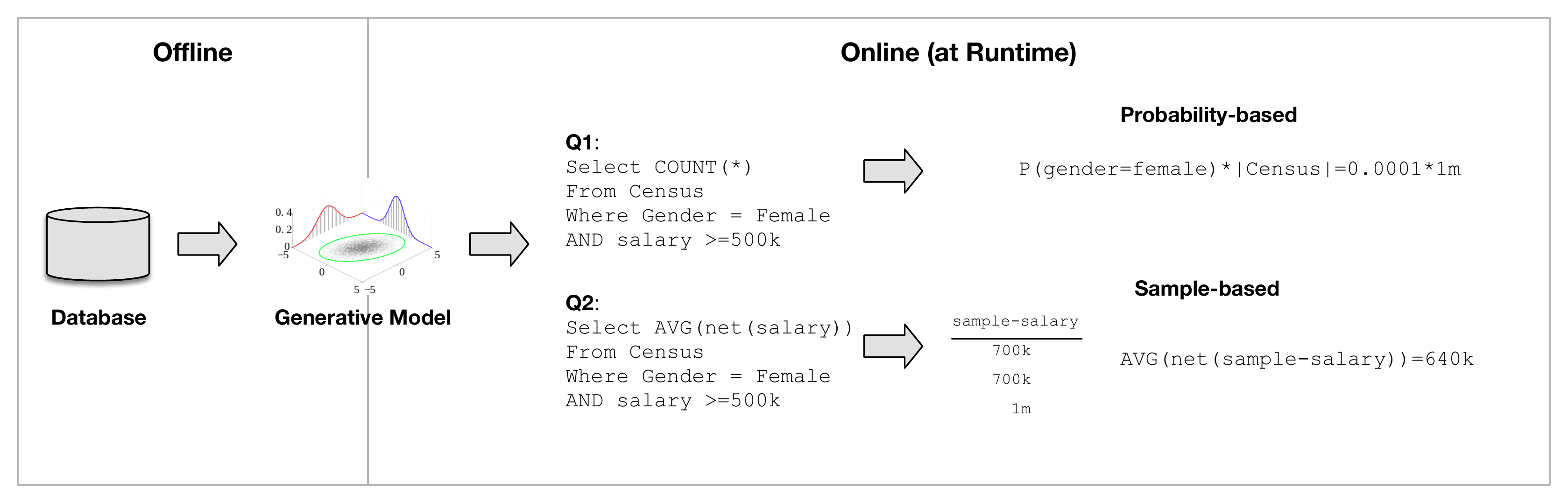}
\caption{Overview of model-based AQP.}
\label{fig:overview}
\end{figure*}

\paragraph{Contribution:} In this paper, we present a new approach to AQP called \emph{Model-based AQP} that leverages generative models learned over a given database to answer SQL queries at interactive speeds. 
Generative models are an unsupervised approach for statistical modeling to learn the joint probability distribution of a given data set.
Different from discriminative models, any attribute of a data set can be used as a target variable for prediction.
To that end, generative models can capture the characteristics over the complete database in a succinct manner without making any prior assumptions. 
Different from other AQP approaches that rely on pre-computed samples or synopses, generative models can deliver high-quality estimates for arbitrary ad-hoc queries even for rare sub-populations.

Therefore, we explain intuitively how generative models can be used to answer SQL queries in an approximate manner.
The main idea is that generative models are either able to directly provide probability estimates that can be used to compute the results of simple aggregate queries or to generate samples for more complex queries that could even include user-defined functions. 
Since generative models capture the joint probability distribution of the complete underlying data set, both these approaches (i.e., probability estimation as well as sample generation) can guarantee good estimates even for rare sub-populations as we will show in our experimental evaluations.
For example, for sample generation, we use the information in the model to generate stratified samples on-the-fly at query time and thus can guarantee that samples are generated even for rare sub-populations.

In summary in this paper we make the following contributions:
(1) To the best of our knowledge, we are the first paper that discusses the possibilities of how generative models can be used for approximate query processing.
(2) We present two different query processing strategies on top of generative models: one based on probability and expectation estimates and one based on sampling.
(3) We analyze the different query processing strategies using an extensive experimental evaluation based on real and synthetic data sets.

We believe that the basic idea of using generative models is not only applicable for AQP, but also represents a more general approach that can also be used for other query-processing related problems including cardinality estimation for query optimization or to build more robust query answering strategies that tolerate data errors (e.g., by generating data for missing values during query processing).

\paragraph{Outline:} 
The remainder of this paper is organized as follows: 
In Section \ref{sec:overview}, we first give an overview of Model-based AQP and discuss the requirements a generative model has to fulfill to be used for AQP. 
Afterward, we explain Mixed Sum-Product-Networks in Section \ref{sec:spn}, a particular class of generative models that satisfies these requirements.
We then show how SQL queries can be compiled into an inference procedure using Sum-Product-Networks in Section \ref{sec:compile} and then explain the two different AQP execution strategies using SPNs in Sections \ref{sec:aqp1} and \ref{sec:aqp2}.
To show the efficiency of Model-based AQP, we present our evaluation results using benchmarks on real and synthetic data sets in Section~\ref{sec:eval}. 
Finally, we discuss related prior work in Section \ref{sec:related} and 
then conclude by discussing planned future extensions in Section \ref{sec:concl}.

%% file: sections/overview.tex
\section{Overview}
\label{sec:overview}

\begin{figure*}[t!]
\centering
\includegraphics[width=1.\textwidth]{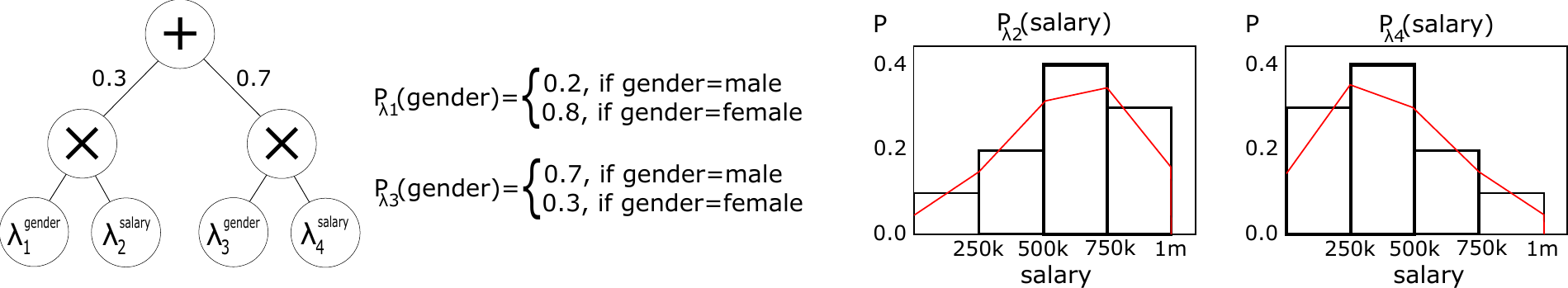}
\caption{Example SPN over variables gender and salary.}
\label{fig:example_spn}
\end{figure*}

\subsection{Model-based AQP}

The main idea of Model-based AQP is shown in Figure \ref{fig:overview}.
The generative model is built once over the original (potentially large) database and then used to answer SQL queries for data exploration in an interactive manner.
The general approach of Model-based AQP is thus similar to classical AQP creating a sample offline, which is used at runtime to answer queries.
However, different from the sampling-based approaches Model-based AQP does not need to know the workload (i.e., queries) to deal with rare sub-populations. 
In this paper, we support aggregate SQL queries with and without filter predicates as well as with and without group-by statements for Model-based AQP.
Furthermore, we support user-defined functions and general arithmetic expressions to be used instead of base attributes.
Joins and nested queries are not covered in this paper, but represent an interesting avenue for future work.

For approximate query processing, Model-based AQP provides different strategies as shown on the right-hand side of Figure~\ref{fig:overview}: a probability-based strategy and a sample-based strategy.
As shown on the right-hand side for query Q1, the probability-based strategy translates the given SQL query directly into an inference procedure and uses the resulting probability as well as statistics (i.e., the size of the table) to answer the query.
The sample-based strategy instead is shown for query Q2.
In this strategy the model is used to generate samples and then the samples are used to answer the query.

Indeed, the probability-based strategy is more efficient than the sample-based strategy but can only be used for simple aggregate queries without user-defined functions or arithmetic expressions and only supports conjunctive predicates (i.e., the net salary in the example).
Moreover, both strategies can deliver estimates efficiently for rare sub-populations (e.g., the females with salary above a $500$k as shown in Figure \ref{fig:overview}) without knowing the workload ahead of time.
In our experimental evaluation, we show that both strategies outperform classical online sampling on skewed data not only in terms of runtime but also in the quality of the approximated results.

In this work, we focus on pure analytical workloads where data is not updated online such as in data warehouses.
We therefore assume that a new SPN can be learned every time a new bulk of data is loaded into the database.
However, this can be improved since no new SPN needs to be learned if the statistical properties of the data after the update do not change. 
In this case, we can reuse the same SPN and only need to update the statistics (i.e., table sizes). 
Detecting this case efficiently is an interesting avenue for future work though.

\subsection{Model Requirements}
\label{sec:model}

An essential requirement for Model-based AQP is that a generative model must enable tractable inference in hybrid domains, consisting of mixed-continuous, discrete and/or categorical distributions to support arbitrary database schemata.
Sum-Product Networks (SPNs) \citep{spn_initial} fulfill the tractability requirements and are therefore are a suitable candidate for AQP. 
In contrast to other probabilistic models, SPNs can efficiently compute a large class of probabilistic queries~\citep{spn_initial}. 

Furthermore, SPNs can generate samples, provide normalized probabilities and expectation estimates with a complexity linear on the size of the model. 
In particular, they are capable of relevance sampling, i.e., they can generate samples for a specific sub-population. 
However, the drawback of SPNs is that the parametric form of the distributions of the features has to be specified in advance~\citep{spn_mspn}. 
This can be quite challenging and time consuming especially for hybrid domains which involve continuous and discrete random variables at the same time~\citep{spn_mspn}.

For that reason, we use so called Mixed-Sum Product Networks (MSPNs)~\cite{spn_mspn}.
MSPNs are a general class of mixed probabilistic models that, by combining Sum-Product Networks and piecewise polynomials, allow for a broad range of exact and tractable inference without making strong distributional assumptions.
Hence, MSPNs require only knowledge about the statistical types of the random variables, fortunately, this is available in the database schema which makes them an ideal candidate for AQP.

%% file: sections/sum_product_networks.tex
\section{Mixed Sum-Product Networks}
\label{sec:spn}

MSPNs (for simplicity we will only use the term SPN in the following) learn the probability distribution $P(X_1,X_2,\dots,X_n)$ of the variables $X_1,\dots,X_n$ which are present in the dataset. For example in Figure \ref{fig:example_spn} the SPN is defined over the discrete variable gender and the continuous variable salary. 

SPNs are rooted acyclic graphs with sum and product nodes as internal nodes and leaves defining probability distributions for single variables \citep{spn_initial}. Intuitively, sum nodes split the population into subgroups and product nodes split independent variables of a population. For example in Figure \ref{fig:example_spn} the top sum node splits the census data into two groups: The left group which is dominated by women and high salaries and the right group with more men and lower salaries. In each of these groups salary and gender are independent and hence split by a product node. The leave nodes determine the probability distributions of the variables gender ($P_{\lambda 1}(\textit{gender})$ and $P_{\lambda 3}(\textit{gender})$) and salary ($P_{\lambda 2}(\textit{salary})$ and $P_{\lambda 4}(\textit{salary}))$ for every group. Linear interpolations (red) of the histograms are used in leaf nodes for continuous variables to approximate the true probability distribution.

The scope of a node is defined as the set of variables occurring in the underlying leaf nodes. For example in Figure \ref{fig:example_spn} the scope of the nodes $\lambda_1$ and $\lambda_3$ is $\{\textit{gender}\}$ and the scope of the product and sum nodes is $\{age, gender\}.$ A SPN representing a valid probability distribution can now be defined recursively ~\citep{spn_mspn}:
(1) A tractable distribution over a single variable is a SPN, 
(2) a product of SPNs which are defined over different scopes is a SPN and 
(3) a sum of SPNs which share the same scope is a SPN. 

In the following, we explain the two basic building blocks of how SPNs can be used for AQP; i.e., (1) how to estimate probabilities and expectations for given sub-populations and (2) how to generate samples for a given sub-population.
Finally, we discuss potential optimizations to enable more efficient AQP.

\subsection{Inference Procedure}

To answer probabilistic queries in a SPN, we evaluate the nodes starting at the leaves. Given some evidence, the probability output of the leaf distributions is propagated bottom up. For product nodes, the values of the children nodes are multiplied and propagated to their parents. For sum nodes, instead, we sum the weighted values of the children nodes. The value at the root indicates the probability of the asked query. To compute marginals, i.e., the probability of partial configurations, we set the probability at the leaves for irrelevant variables to one and then proceed as before. Especially when dealing with SQL queries, not all variables are usually of interest for answering a query. Conditional probabilities can then be computed as the ratio of partial configurations. 

Assume we want to estimate the probability that a member of the census data is a women earning more than 500k. The respective conditions would be $C_{\textit{gender}}=\{\textit{female}\}$ and $C_{\textit{salary}}=\{[500k,1m]\}.$ In order to estimate the probability $P(C_{\textit{gender}}\land C_{\textit{salary}})$ we first apply the gender condition to all leave nodes with scope $\{\textit{gender}\}.$ The probabilities for the discrete random variable gender are simply $P_{\lambda 1}(\textit{gender=female})=0.8$ and \linebreak $P_{\lambda 3}(\textit{gender=female})=0.3.$ For the continuous variable, the probabilities are approximated with the linear interpolations of the histograms. Let us assume the probabilities are $P_{\lambda 2}(500k \le \textit{salary} \le 1m)=0.7$ and $P_{\lambda 4}(500k \le \textit{salary} \le 1m)=0.3.$ In the bottom-up pass we obtain $P(C_{\textit{gender}}\land C_{\textit{salary}})=0.231$. 

Another essential building block for this work is the ability of SPNs to compute expectations $E[X_i |C_1\land\dots\land C_n]$ for a variable $X_i$ and given conditions $C_1$ to $C_n.$ Similar to the computation of the probability, the expectation is computed in a bottom-up process starting at the leaves of the SPN. For all leaves with scope~$\{X_i\}$, the expectation of the respective distribution is evaluated while for all other variables inference is computed according to the conditions. In the case of computing the expectation on a sub-population of the data, the result obtained at the root node needs to be normalized according to the probability $P(C_1\land \ldots \land C_n)$.

\subsection{Sample Creation}

We first explain how random sampling without any conditions works. 
Let $|S|$ denote the number of samples that should be generated with the help of the SPN. The intuition of sum nodes is that they describe different sub-populations. Hence, every child of a sum node has to generate samples proportional to the weight. For a sample size $|S|$ of 100 in Figure \ref{fig:example_spn}, approximately 30 samples would be generated by the left sub-SPN and 70 samples by the right sub-SPN. The children of product nodes have disjoint scopes. Hence, one can sample from the children independently and concatenate the results of the different scopes. For example, for the left sub-SPN a gender would be drawn from $P_{\lambda 1}$ and a salary from $P_{\lambda 3}$. Both values would be combined to constitute one of the 30 samples of the left sub-SPN. 

The most valuable property of SPNs for AQP is the ability to generate biased samples, i.e. samples that satisfy conditions $C_1$ to $C_n$. This requires an additional initialization step. In this step, the weights of the sum nodes are adjusted. For every child node, the weight is multiplied with the probability $P(C_1\land \dots \land C_n)$ computed on the sub-SPN of the child node. Afterwards, the weights of the sum nodes have to be normalized so that they sum up to one. For example, if only women should be sampled from the SPN of Figure \ref{fig:example_spn} the left weight of the sum node becomes $0.3*P_{\lambda 1}(gender=female)=0.3*0.8=0.24$ and the right weight $0.7*P_{\lambda 3}(gender=female)=0.7*0.3=0.21.$ As the weights do not sum up to one they have to be scaled resulting in a left weight of $0.53$ and a right weight of $0.47.$ After this step, we can use the same algorithm as for random sampling with the only difference that leaf nodes are restricted to values satisfying the conditions.

\subsection{Optimizations for AQP}

Creating samples for continuous random variables in a SPN relies on rejection sampling \citep{spn_mspn}.
However, rejection sampling is known to be computationally inefficient since many sample candidates need to be created before a candidate is accepted.
For that reason, we modified the basic structure of SPNs to store already a materialized sample of data in the leave nodes for continuous attributes instead of the distribution.
That way, sampling can use the SPN as an index and directly use the original data for sample creation as we will discuss in Section \ref{sec:aqp2}.

Another way to improve the efficiency of SPNs is to marginalize the SPN to the set of variables which are relevant for the query. We simply cut off all the child nodes of the SPN that contain irrelevant variables. In particular, if the query only operates on a few variables, the size of the SPN can be reduced significantly. Especially for sampling this is very useful because we avoid generating values for irrelevant variables. Moreover, after cutting off the irrelevant child nodes, the structure of the SPN can be further collapsed by combining sum nodes or removing product nodes.

%% file: sections/query_compilation.tex
\section{Query Compilation}
\label{sec:compile}

In the following, we explain how the filter $E$ and the grouping $G$ of an aggregation query as shown in Listing~\ref{lst:basic_sql_query} can be compiled into conditions for an SPN (i.e., to compute a probability or create samples as discussed before).

\begin{lstlisting}[language=SQL,escapechar=@,language=SQL,basicstyle=\ttfamily, caption={Basic SQL-query with an aggregation.}, label={lst:basic_sql_query}]
SELECT G, AGGR(A) FROM T WHERE E GROUP BY G
\end{lstlisting}

The details on how the conditions are used to compute the actual query result in an approximate manner will be discussed in Sections \ref{sec:aqp1} to \ref{sec:aqp2}.

\subsection{Filter Predicates}

The filter predicate $E$ defines the sub-population of the data which the SQL query needs.
The compilation of a simple predicate of the form \emph{att OP const} where \emph{att} is an attribute of the table T, \emph{OP} is one of the operators $\{=,\neq, <, \leq, >, \geq\}$  and \emph{const} is a constant from the domain of \emph{att} is straightforward by transforming the predicate into a set of possible values for discrete random variables or a set of value ranges for continuous random variables. 
However, the predicate $E$ can contain a conjunctive or disjunctive combination of conditions on same or different columns. Since we are only able to process conjunctive conditions with a SPN, these combinations need to be handled differently which we examine in the following sections.

In the case of conjunctive conditions, we extract the set of possible values or value ranges for each particular condition independently. If multiple conditions are applied to the same column, the intersection of the extracted sets is used. As a result, we obtain a condition for each particular column of the filter which can then be used by a SPN.

When dealing with disjunctive conditions, we have to consider two cases: (1)~The conditions are either applied to the same column or~(2) the conditions are applied to different columns. In the first case, we take the union of the extracted sets. However, the second case needs to be handled in a particular manner because an SPN is not able to process queries with disjunctive conditions on different random variables. 
Therefore, the disjunctive condition needs to be transformed into multiple conjunctive conditions by using the addition rule.

\subsection{Group-by Attributes}

In order to translate a grouping condition, we need to know a distinct list of group-by values $g \in G$.
The distinct list of group-by values $g$ can be efficiently derived from the SPN directly.
For each particular group-by value $g$, a separate query needs to be computed on each of the specified sub-populations of the data. The result of the group-by is then the union of all queries.

%% file: sections/probability_based.tex
\section{Strategy I: Probability-based}
\label{sec:aqp1}

First, we want to introduce a query execution strategy that relies on expectation and probability estimations using the SPN as discussed in Section \ref{sec:compile}.
This strategy is preferred since no samples have to be drawn and thus it is the most efficient strategy. 
However, it is only applicable to simple aggregate queries with no user-defined functions and exclusively conjunctive predicates.

As already stated in the previous section, a given SQL query can be parsed to obtain conditions corresponding to the filter predicate $E$ and the group-by attributes $G$. 
Depending on the aggregation function, we have to perform different computations with the SPN. In the following, we discuss the details of the computation for the aggregation functions \emph{COUNT}, \emph{AVG} and \emph{SUM}.
Table \ref{tab:measure_notation} summarizes the results.
\emph{MIN} and \emph{MAX} as well as other aggregation functions are currently not supported similar to other AQP approaches.

\subsection{COUNT}

In order to provide an answer for a query with the aggregation function \emph{COUNT} we have to determine the number of entries of the queried sub-population defined by $E$. This can be estimated by multiplying the table size $|T|$ with the probability for the sub-population $P(E)$. Moreover, in case of a grouping we additionally have to analyze this sub-population with respect to the individual groups $g$ defined by $G$. 

Equation~\ref{eq:model_based:count} shows how the results for each particular group $g \in G$ are computed. The probability $P(E \land g)$ represents that a record fulfills the filter condition $E$ and is in the group $g$ is multiplied with the table size. In case that $E$ and $g$ share conditions on same columns, these conditions are combined by taking the intersection of the set of possible values or value ranges.

\begin{equation}
    \label{eq:model_based:count}
    P(E \land g) * |T|
\end{equation}

In case the query does not specify a group-by statement, the computation simplifies to $P(E) *|T|$.

\subsection{AVG}

For the computation of an \emph{AVG} aggregation, we rely on the functionality of SPNs to compute expectations. Like for the estimation of a \emph{COUNT} aggregation we first have to parse the SQL-query to obtain the conditions for the sub-population $E$ and the grouping $G$. In addition, we need to extract the columns of $A$ on which the aggregation is applied. In case that $A$ is an arithmetic expression of columns, a result for the query can only be computed if only the operators for addition and subtraction are used. With respect to these operators, we first compute the expectation for every single column of $A$ individually which are then added or subtracted afterwards. In contrast, arithmetic expressions with multiplication and division operators cannot be computed with the probability-based approach. The computation for each particular group $g \in  G$ according to the sub-population $E$ is displayed in Equation~\ref{eq:model_based:avg}.

\begin{equation}
    \label{eq:model_based:avg}
    E(A | E \land g)
\end{equation}

\subsection{SUM}

The computation of the result for a \emph{SUM} aggregation can be reduced to the computation of a \emph{COUNT} and an \emph{AVG} aggregation by multiplying the respective results. Equation~\ref{eq:model_based:sum} represents the computation of a \emph{SUM} aggregation based on the queried sub-population $E$ and the grouping $G$. For the same reason as for the \emph{AVG} aggregation, the \emph{SUM} aggregation can only be applied on single columns and on arithmetic expression of columns which use the operators addition and subtraction. In any other case, no result with the probability-based approach can be computed and the sample-based approach, which is discussed next, will be used.

\begin{equation}
    \label{eq:model_based:sum}
    E(A | E \land g) * P(E \land g) * |T|
\end{equation}

\begin{table}[!t]
\centering
\caption{Aggregations for probability-based AQP.}
\label{tab:measure_notation}
\footnotesize
\begin{tabular}{l l}
	\toprule
	\textbf{Aggregation} & \textbf{Computation for each group $g\in G$}   \\
\midrule
COUNT 			& $P(E \land g) * |T|$\\
AVG             & $E(A | E \land g)$ \\
SUM             & $E(A | E \land g) * P(E \land g) * |T| $\\
\bottomrule
\end{tabular}
\end{table}

%% file: sections/sample_based.tex
\section{Strategy II: Sample-based}
\label{sec:aqp2}

In addition to the probability-based approach, the ability to generate samples with a SPN offers us another way to approximate the result for more complex aggregation queries which use user-defined functions, arithmetic expressions or disjunctive predicates. 
Contrary to classical sample-based approaches for AQP, we can produce biased samples online at query runtime with the SPN, which is one of the significant advantages of SPNs.
In this work we propose three different sampling techniques which are explained in the following sections.

\begin{figure*}
\scriptsize
\begin{align*}
\hspace{0.42cm} P(\textit{filter} = 1) &= 25\% & P(\textit{A}= 1 | \textit{filter} = 1) &= 20\% & P(\textit{A}= 1 | \textit{filter} = 2) &= 70\% & P(\textit{A} = 1 | \textit{filter} = 3) &= 90\% & P(\textit{A}= 1 | \textit{filter} = 4) &=  99\% & P(\textit{B} | \textit{A} = 1) &= N(\mu=100, \sigma=20)\\
P(\textit{filter} = 2) &= 25\% & P(\textit{A}= 2 | \textit{filter} = 1) &= 20\% & P(\textit{A}= 2 | \textit{filter} = 2) &= 20\% & P(\textit{A} = 2 | \textit{filter} = 3) &=   6\% & P(\textit{A}= 2 | \textit{filter} = 4) &= 0.6\% & P(\textit{B} | \textit{A} = 2) &= N(\mu=110, \sigma=20)\\
P(\textit{filter} = 3) &= 25\% & P(\textit{A}= 3 | \textit{filter} = 1) &= 20\% & P(\textit{A}= 3 | \textit{filter} = 2) &=   6\% & P(\textit{A} = 3 | \textit{filter} = 3) &=   3\% & P(\textit{A}= 3 | \textit{filter} = 4) &= 0.3\% & P(\textit{B} | \textit{A} = 3) &= N(\mu=120, \sigma=20)\\
P(\textit{filter} = 4) &= 25\% & P(\textit{A}= 4 | \textit{filter} = 1) &= 20\% & P(\textit{A}= 4 | \textit{filter} = 2) &=   3\% & P(\textit{A} = 4 | \textit{filter} = 3) &=0.9\% & P(\textit{A}= 4 | \textit{filter} = 4) &= 0.09\% & P(\textit{B} | \textit{A} = 4) &= N(\mu=130, \sigma=20)\\
    & & P(\textit{A}= 5 | \textit{filter} = 1) &= 20\% & P(\textit{A}= 5 | \textit{filter} = 2) &=   1\% & P(\textit{A} = 5 | \textit{filter} = 3) &=0.1\% & P(\textit{A}= 5 | \textit{filter} = 4) &= 0.01\% & P(\textit{B} | \textit{A} = 5) &= N(\mu=140, \sigma=20)
\end{align*}
\caption{Probability distributions of the synthetic dataset.}
\label{fig:probability_distribution_synthetic}
\end{figure*}

\subsection{Random Sampling}

First, as a baseline we introduce the generation of random samples with the SPN for which we use the sample functionality of the SPN without specifying any conditions. Like for classical random sampling from data, we face the issue that samples can be generated which are not relevant for answering the submitted query. In particular, if the SQL query is only applied to a small sub-population of the data, many generated samples are discarded.

Using random samples, the query result can be approximated. In case of an \emph{AVG} aggregation, no modifications have to be made since the result of an \emph{AVG} aggregation is independent of the number of entries on which it is computed. In contrast, \emph{COUNT} and \emph{SUM} aggregations depend on the number of samples on which they are computed. Therefore, we have to scale-up the result of these aggregations to get the approximation. This is done by multiplying the result with the total number of entries of the data divided by the number of the samples which have been generated. 
We refer to this multiplier as $m_{random}$ and it is given in Equation~\ref{eq:sample_based:rel_random}.

\begin{equation}
    \label{eq:sample_based:rel_random}
    m_{random} = \frac{|T|}{|S|}
\end{equation}

\subsection{Relevance Sampling}

In order to avoid the generation of irrelevant samples, we use another more advanced approach called relevance sampling. This approach only generates samples for the queried sub-population defined by the filter $E$. Compared to the random sampling approach, we can improve the efficiency of approximating the result, especially for rare sub-populations, since we do not have to discard any samples. For example, if only $1\%$ of the data is relevant for the SQL query, then the relevance sampling approach is one hundred times more efficient than the random sampling approach to obtain the same precision for the approximation.

Due to the sample generation, the approximation of the result for the aggregation is different compared to random sampling. To scale-up the result of \emph{COUNT} and \emph{SUM} aggregation queries, we further need to multiply the result with the probability $P(E)$ of the sub-population of the data. This probability is obtained by performing inference for the sub-population specified by the filter $E$. 
The respective multiplier used to scale-up the result is computed as $m_{relevance}$ in Equation~\ref{eq:sample_based:rel_relevance}.

\begin{equation}
    \label{eq:sample_based:rel_relevance}
    m_{relevance} = \frac{|T|}{|S|} P(E)
\end{equation}

\subsection{Stratified Sampling}

The relevance sampling approach is already a major improvement for the approximation of aggregation results compared to random sampling but it ignores the grouping of the SQL query. Each specific queried group should obtain an approximation as fast as possible. However, the relevance method does not consider the selectivity of the queried groups which can be skewed. Hence, particular groups will obtain more samples to approximate the result than other groups which has an effect on the precision of the approximations. 

This problem can be solved with the stratified sampling approach which can generate samples for each particular group~$g$ independently. We introduce the notation $|S(g)|$ representing the number of samples which need to be generated for a group~$g$. In general, the total number of samples to generate is evenly distributed among the groups. However, in order to avoid over-representation of rare sub-populations we restrict to generate more samples for a group than samples are available for that group in the original data. This information can be obtained by performing inference on the SPN with the conditions for that particular group. The remaining number of samples, which have been cut off, are distributed evenly over the other groups, for which the same restriction is applied.

Similar to relevance sampling, we have to adapt the computation of the approximation, because we rely on biased sampling. Since we are generating the samples for each group independently, the aggregation result has to be approximated for each group on its own.
The multiplier $m_{stratified}(g)$ for each particular group $g$ is given in Equation~\ref{eq:sample_based:rel_stratified}. Similar to the other proposed sampling approaches, the multiplier only needs to be applied to scale-up the result of the \emph{COUNT} and \emph{SUM} aggregations.

\begin{equation}
    \label{eq:sample_based:rel_stratified}
    m_{stratified}(g) = \frac{|T|}{|S(g)|} P(E \land g)
\end{equation}

%% file: sections/experimental_evaluation.tex
\begin{figure*}[t!]
\includegraphics[width=\linewidth]{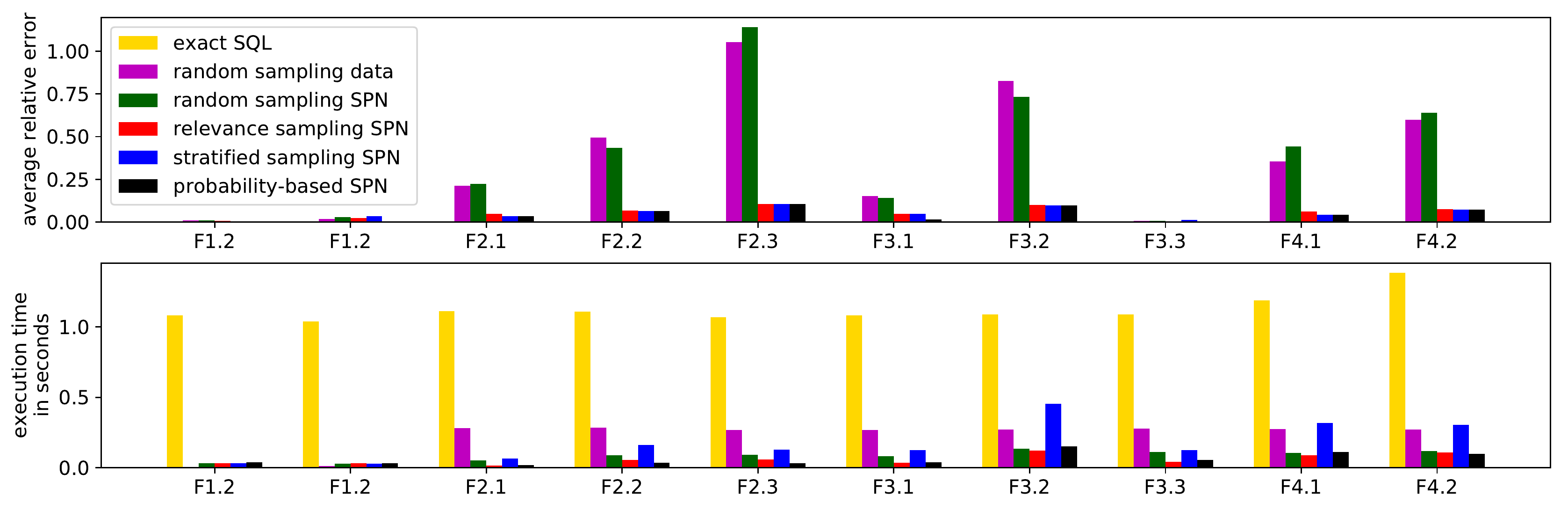}
\caption{Results of the average relative error and the execution time for all queries on dataset \emph{flights10M}.}
\label{fig:evaluation:comparision_error}
\end{figure*}

\section{Experimental Evaluation}
\label{sec:eval}

A key aspect of our experimental evaluation is to demonstrate that our proposed AQP techniques based on SPNs are superior to random sampling from data. Therefore, we have first evaluated the efficiency of our approaches on a real dataset with different sizes. Furthermore, we executed additional experiments on synthetic data to show how our proposed approaches behave by varying the selectivity of particular SQL queries. Similarly, we have evaluated the effect of skewness of the group-by attributes on the approximations as well as how the quality of the learned model with respect to different parameter settings influences query results.

In the following we first explain the experimental setup (metrics, data sets, queries as well as other important settings) before we discuss the results of our experimental evaluations.

\begin{table}[!t]
\centering
\footnotesize
\caption{Statistics about the datasets.}
\label{tab:evaluation:dataset_statistics}

\begin{tabular}{lrrrr}
\toprule
		 \textbf{name}  & \textbf{\#instances} & \multicolumn{1}{c}{\textbf{size in}} & \multicolumn{1}{c}{\textbf{\#discrete}} & \multicolumn{1}{c}{\textbf{\#continuous}}  \\
		                & & \multicolumn{1}{c}{\textbf{megabyte}} & \multicolumn{1}{c}{\textbf{columns}} & \multicolumn{1}{c}{\textbf{columns}} \\
\midrule
flights100K &     100,000 &  5.55& 6 &        6 \\
flights1M &    1,000,000 &   55.52& 6 &        6 \\
flights10M &   10,000,000 &  555.21& 6 &        6 \\
synthetic &    1,000,000 &   23.29& 2 &        1 \\
\bottomrule
\end{tabular}
\end{table}

\begin{table}[!t]
\centering
\footnotesize
\caption{Statistics about the queries.}
\label{tab:evaluation:flight_queries}

\begin{tabular}{llrrr}
\toprule
\textbf{identifier} & \textbf{aggregation} & \text{groups} & \textbf{selectivity (in \%)} & \textbf{skewness}\\
\midrule
     F 1.1 &                AVG &                 1   &      $\sim5.6000$  &       - \\
     F 1.2 &                AVG &                 1   &       $\sim1.3800$ &       - \\
     \midrule
     F 2.1 &                COUNT &                 26 &       $\sim1.0000$ &    1.4343 \\
     F 2.2 &                COUNT &                 26&        $\sim0.1260$ &    1.4122 \\
     F 2.3 &                COUNT &                 22&       $\sim0.0140$ &    0.4719 \\
     \midrule
     F 3.1 &                SUM &                 22&       $\sim1.0000$ &    0.0329 \\
     F 3.2 &               SUM &                 53&       $\sim0.1200$ &    2.4420 \\
     F 3.3 &                AVG &                 26&       $\sim0.1500$ &    1.4735 \\
     \midrule
     F 4.1 &                COUNT &                53&       $\sim1.3600$ &    2.4234 \\
     F 4.2 &               COUNT &                26&       $\sim0.1000$ &    1.5480 \\
     \midrule
    S 1.1 &                 COUNT &                5 &      $\sim25.0000$ &   -0.3851 \\
    S 1.2 &                 COUNT &                 5 &      $\sim25.0000$ &    1.2751 \\
    S 1.3 &                 COUNT &                 5 &      $\sim25.0000$ &    1.4876 \\
    S 1.4 &                 COUNT &                 5 &      $\sim25.0000$ &    1.4999 \\
\midrule
    S 2.1 &                 AVG &                  5 &      $\sim25.0000$ &   -0.3851 \\
    S 2.2 &                   AVG &                  5 &      $\sim25.0000$ &    1.2751 \\
    S 2.3 &                  AVG &                  5 &      $\sim25.0000$ &    1.4876 \\
    S 2.4 &                  AVG &                  5 &      $\sim25.0000$ &    1.4999 \\
\midrule
    S 3.1 &                 COUNT &       1 & $\sim5.0000$  & -\\
    S 3.2 &                 COUNT &      1 &$\sim0.2500$   & - \\
    S 3.3 &                 COUNT &      1 &$\sim0.0250$  & - \\
    S 3.4 &                 COUNT &      1 & $\sim0.0025$ & - \\
\midrule
    S 4.1 &                 AVG &       1 &$\sim5.0000$  & - \\
    S 4.2 &                  AVG &       1 &$\sim0.2500$  & - \\
    S 4.3 &                  AVG &      1 &$\sim0.0250$   & - \\
    S 4.4 &                   AVG &       1 &$\sim0.0025$  & - \\
\bottomrule
\end{tabular}
\end{table}

\subsection{Setup and Workload}

\paragraph{Metrics:} 
For reporting the metrics, we have repeated each experiment ten times and averaged the results. Besides measuring the time to provide an estimate for an aggregation query, we use two evaluation measures in order to evaluate the quality of approximation results which we have taken from~\cite{bide}. We denote $G$ as the actual result of an aggregation which is commonly known as the \emph{ground truth}. Furthermore, we define $G_{groups}$ as a set which contains all the groups which appear in the results. In order to access the result for a specific group $i$ we denote $G_{values}(i)$. Similar to this notation we represent the result for the approximation of an aggregation as $A$, $A_{groups}$ and $A_{values}(i)$.

First of all, an approximation algorithm is not always able to provide approximations for all groups which appear in the \emph{ground truth}. In order to measure the coverage of the groups in the approximation, we use the measure \emph{bin completeness} which is defined in Equation~\ref{measure:missing_bins}.

\begin{equation}
	\label{measure:missing_bins}
	\text{bin completeness} = \frac{|G_{groups} - A_{groups}|}{|G_{groups}|}
\end{equation}

On the other hand, we examine the quality of the approximation by using the relative error which represents the relative difference between the approximated value and the actual value for a particular group. The relative error for a group $i$ can only be computed if $G_{values}(i)$ and $A_{values}(i)$ exist, therefore, it can only be applied on the set $X = G_{groups} \cap A_{groups}$. In order to obtain a measure for the whole query, we average the relative error for all groups resulting in the average relative error which is displayed in Equation~\ref{measure:avg_rel_error}.

\begin{equation}
	\label{measure:avg_rel_error}
	\text{avg. relative error} = \frac{1}{|X|} \sum_{i \in X}  \frac{|G_{values}(i) - A_{values}(i)|}{|G_{values}(i)|}
\end{equation}

\paragraph{Datasets:} On the one hand, we evaluated our proposed AQP techniques on a real-world dataset (called \emph{flights} database) which we have taken from~\cite{bide}. This dataset represents basic information about particular flights which have been tracked over a couple of years. In order to evaluate different sizes of the dataset we have used the data generator proposed by~\cite{bide} which has the ability to scale a given dataset to a specific size. On the other hand, we have generated a synthetic dataset in order to directly measure the effect of selectivity and skewness on the quality of approximation results when using our approach compared to the other baselines. For that purpose, we have created a dataset with three columns, named \emph{A} (used for grouping), \emph{B} (used for aggregation) and \emph{filter} (used for the predicate). Depending on the filter condition, the resulting distribution of the group-by attributes is either uniform or skewed to a certain degree. Furthermore, correlations between the columns \emph{A} and \emph{B} exist. The underlying probability distribution of the \emph{synthetic} dataset is shown in Figure~\ref{fig:probability_distribution_synthetic}. Basic statistics of the two datasets can be found in Table~\ref{tab:evaluation:dataset_statistics}. We decided not to use standard synthetic benchmarks like TPC-H since they are neither real data nor do they allow to change different parameters such as skeweness and correlation which we wanted to vary in our experiments.

\paragraph{Queries:} In order to evaluate our approaches on a wide variety of different queries, we provide ten queries for the \emph{flights} dataset and twelve queries for the \emph{synthetic} dataset which can be found in the appendix. These queries mainly differ in the number and types of columns they are applied on, the aggregation function, the selectivity of the queried sub-population and the skewness of the grouping. A particular focus of our evaluation is based on queries with a very low selectivity. Basic statistics about the queries are displayed in Table~\ref{tab:evaluation:flight_queries}. The skewness for the grouping is computed with Equation~\ref{eq:skew} in which $Y$ contains the selectivity for each group of the query.

\begin{equation}
\label{eq:skew}
skewness(Y) = \frac{\sum_{i=1}^{|Y|}(Y_{i} - \bar{Y})^{3}/|Y|} {std(Y)^{3}}
\end{equation}

\paragraph{Implementation:} In order to show the effects of our Model-based approaches compared to classical query processing approach-es in terms of runtime and quality, we have implemented all prototypes in a single-threaded query execution engine.
As a first baseline, we have implemented a query execution engine that can execute SQL queries in an exact manner using in-memory arrays to store the database. 
Additionally, we used the same engine but implemented a simple random sampling procedure and used online aggregation to compute approximate query results as described in \cite{aqp_online_aggregation_base}. 
For the evaluation of our Model-based AQP techniques, we loaded the SPNs into memory before execution. 
Hence, all mentioned algorithms maintain their respective data in memory. 

\begin{table*}[!htb]
\centering
\caption{Average relative error of the probability-based approach on different sizes of the dataset \emph{flights}.}
\label{tab:scalability:error}
\footnotesize
\begin{tabular}{rrrrrrrrrrr}
\toprule
 \textbf{size of dataset} &        \textbf{F 1.1} &        \textbf{F 1.2} &        \textbf{F 2.1} &        \textbf{F 2.2} &        \textbf{F 2.3} &        \textbf{F 3.1} &        \textbf{F 3.2} &        \textbf{F 3.3} &        \textbf{F 4.1} &       \textbf{F 4.2} \\
\midrule
$100,000$ &  0.000757 &  0.008604 &  0.249744 &  0.456241 &  0.415543 &  0.090410 &  0.437463 &  0.006803 &  0.269545 &  0.728370 \\
$1,000,000$ &  0.000315 &  0.002836 &  0.081535 &  0.263019 &  0.548479 &  0.043265 &  0.322761 &  0.003539 &  0.126056 &  0.156574 \\
$10,000,000$ &  0.000026 &  0.003425 &  0.033385 &  0.064262 &  0.105324 &  0.016804 &  0.097058 &  0.002170 &  0.043148 &  0.072445 \\
\bottomrule
\end{tabular}
\end{table*}

\begin{figure*}[!t]
\centering
\begin{subfigure}{0.245\textwidth}
\centering
\includegraphics[width=\linewidth]{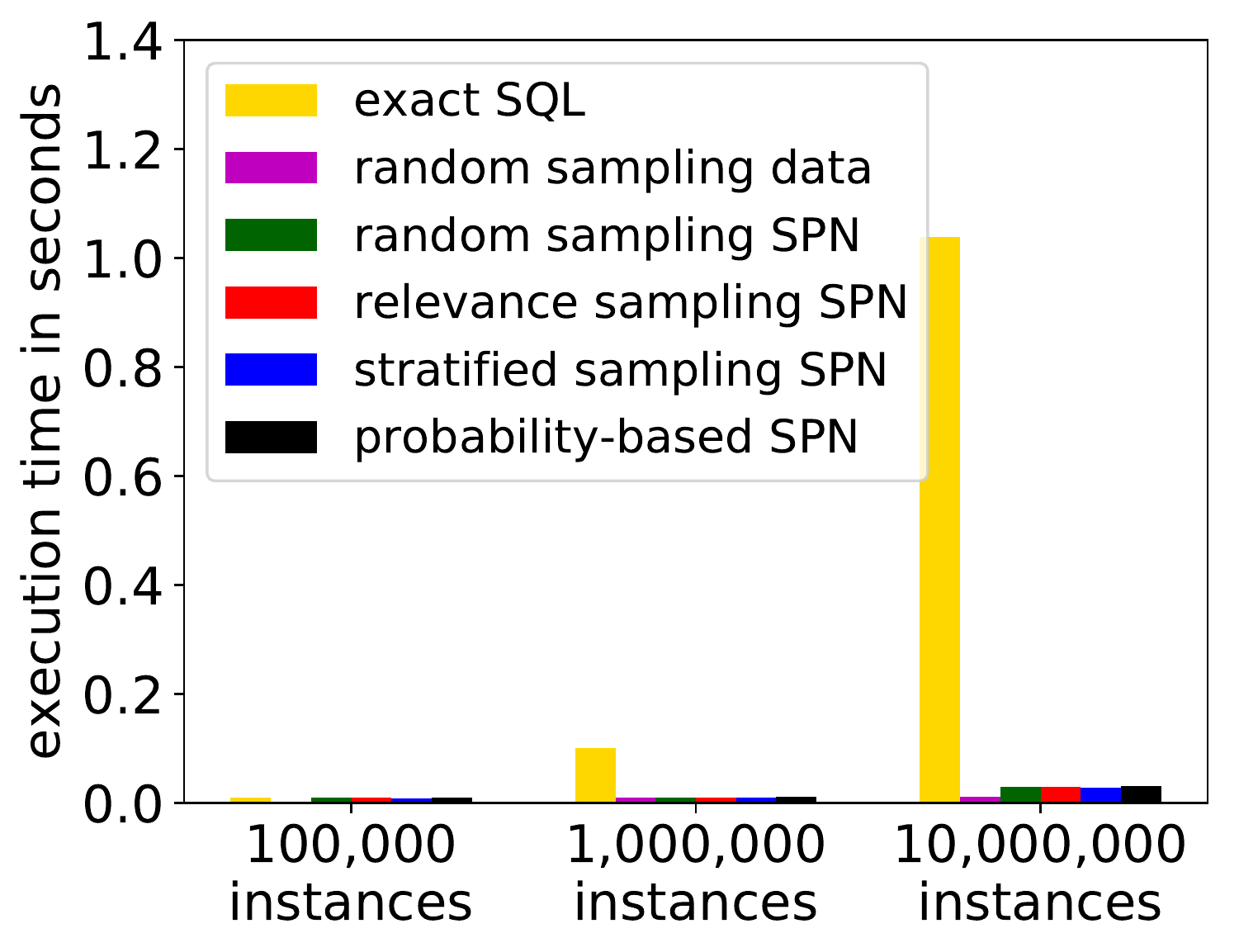}
\caption{F1.2}
\end{subfigure}
\begin{subfigure}{0.245\textwidth}
\centering
\includegraphics[width=\linewidth]{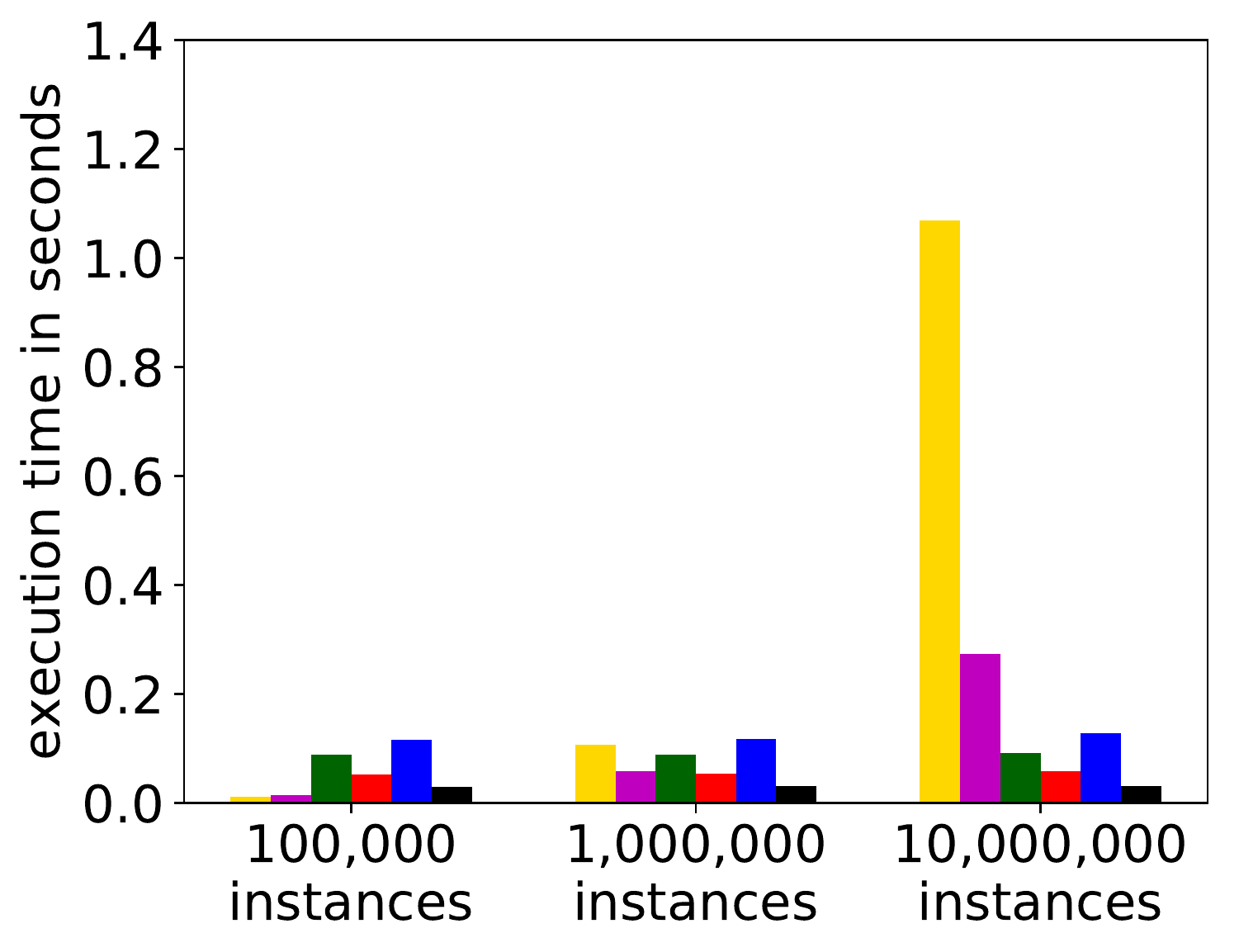}
\caption{F2.3}
\end{subfigure}
\begin{subfigure}{0.245\textwidth}
\centering
\includegraphics[width=\linewidth]{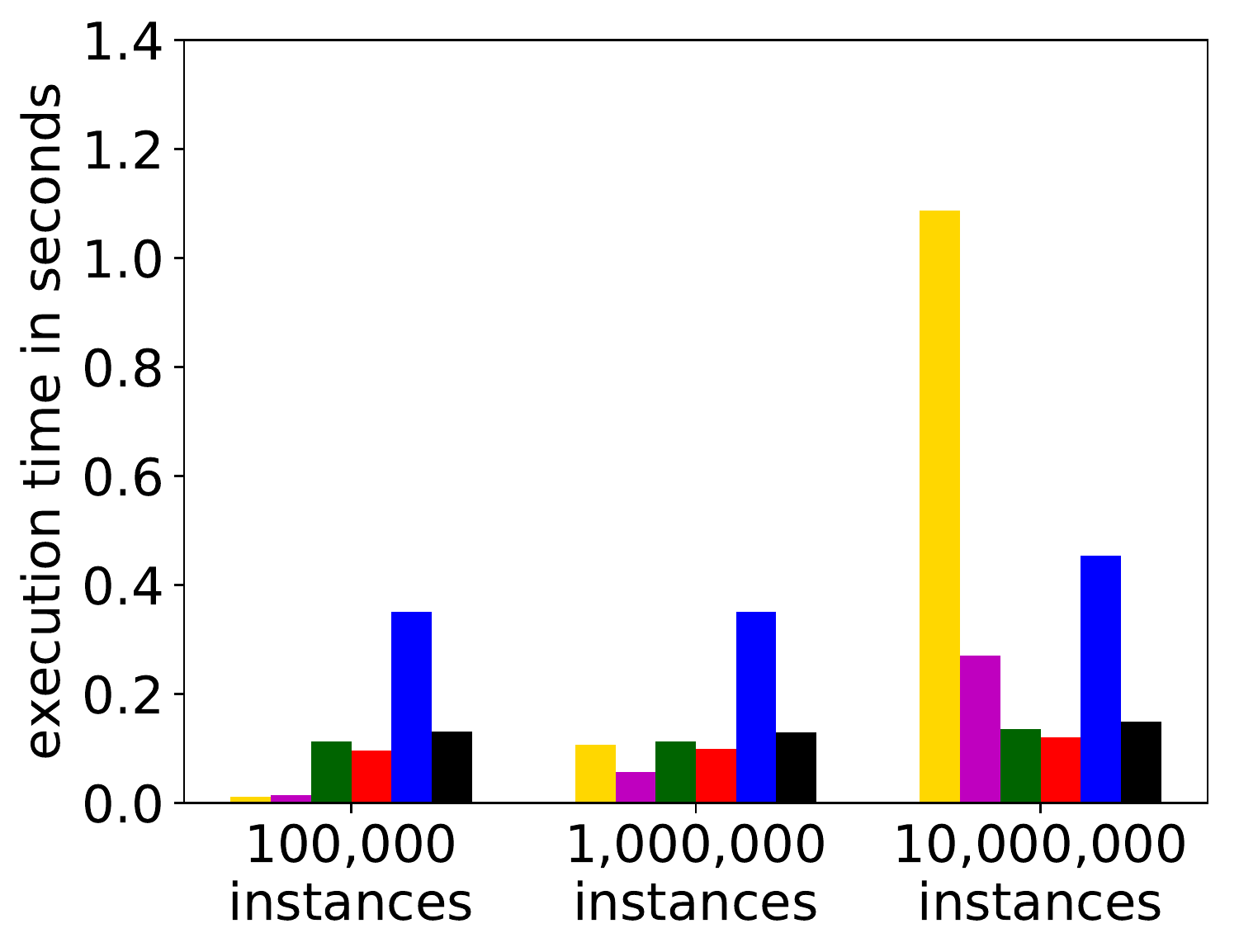}
\caption{F3.2}
\end{subfigure}
\begin{subfigure}{0.245\textwidth}
\centering
\includegraphics[width=\linewidth]{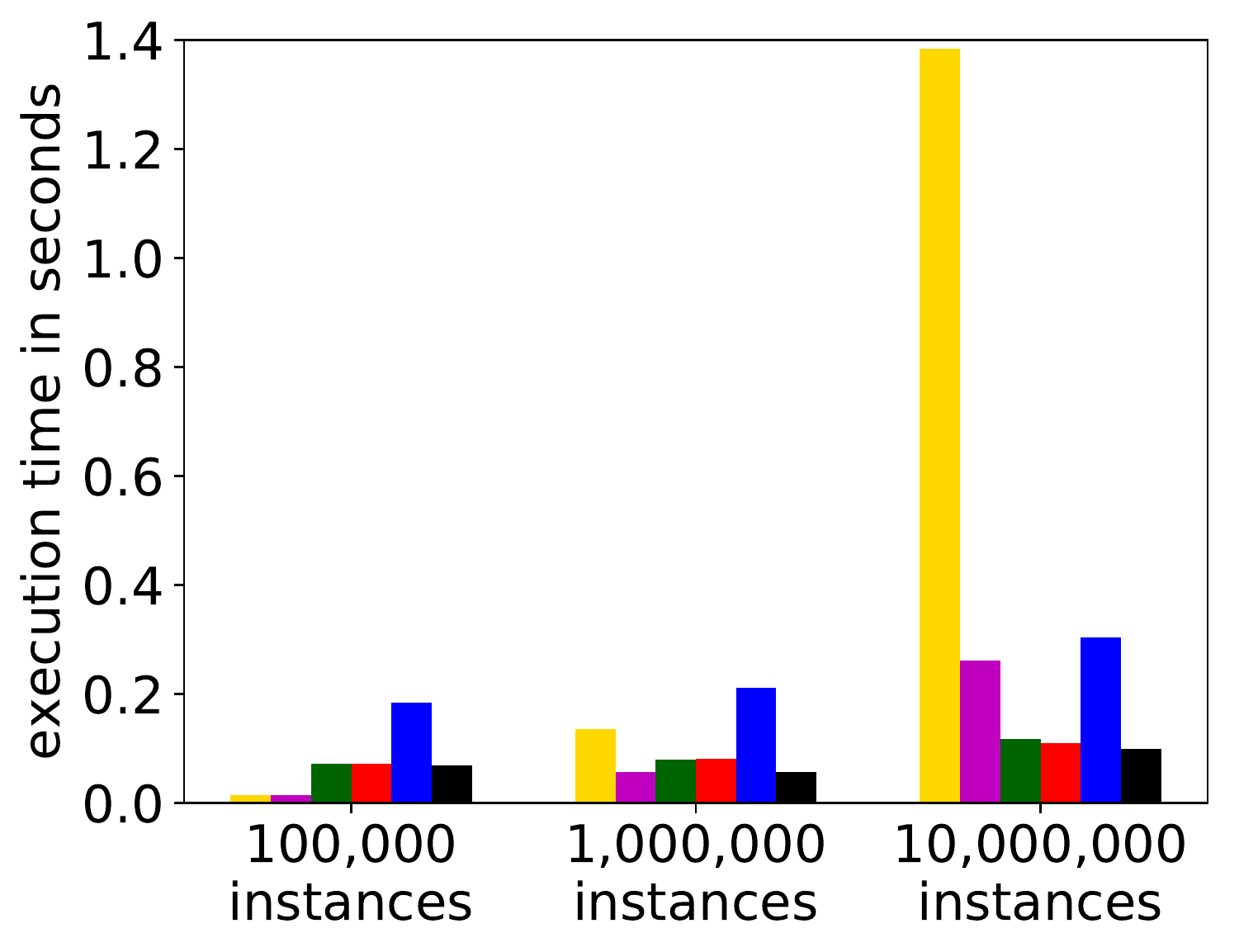}
\caption{F4.2}
\end{subfigure}
\caption{Runtime of the approaches on different sizes of the dataset \emph{flights}.}
\label{fig:scalability:bar_plot}
\end{figure*}

\subsection{Exp. 1: Overall Efficiency}

One major aspect of our experimental evaluation is to compare our proposed AQP techniques with random sampling from data and exact SQL. For that purpose, we have evaluated all AQP approaches on the dataset \emph{flights10M}. In particular, for the sample-based approaches, the idea of this experiment is to report the execution time of the queries until the average relative error is below $5$\% and full bin completeness is achieved. Therefore, we generate samples until the specified goal or the limit of $100,000$ instances is reached. In case that the limit is reached, we stop the sampling procedure and report the error of the results which is achieved with the respective number of samples. Since exact SQL and the probability-based approach do not rely on samples, we only report the average relative error and the execution time for these approaches. 

The results for the average relative error and the execution time for all evaluated queries are visualized in Figure~\ref{fig:evaluation:comparision_error}.
Our first and foremost observation is that our proposed approaches are able to process the queries with typically less computation time comapred to exact SQL and better accuracy compared to random sampling. 
Furthermore, we could observe that the execution time of the stratified sampling approach and the probability-based approach depends on the number of groups in a query. In particular, we found out that the execution time of these approaches for the queries \emph{F3.2} and \emph{F4.1} is higher than for the other queries. The reason for this is that each group needs to be handled individually by these approaches. 

In addition, we detect a minor increase of the runtime in the case that the query contains conditions on continuous columns (e.g. query \emph{F4.2} contains two continuous columns). This is due to the fact that for continuous columns more values need to be accessed in the leaves to process the query. On the other hand, for discrete columns the values in the leaves can be represented compactly by frequency distributions and accessed quickly.

Moreover, if we compare the execution time of the random sampling approaches we can see that for the creation of many samples the sampling from the SPN is much more efficient than from data while the quality of the approximations is similar. By using the SPN, we avoid dealing with huge amounts of data and we only generate samples which are relevant to answer the query which gives us a huge gain in efficiency. This suggests that, in particular for big datasets, relevance sampling from the SPN can be a better choice than random sampling from data. 

According to the quality of the approximations, we can further observe that relevance sampling, stratified sampling and our probability-based approach outperform random sampling on almost all queries. In particular on queries with a very low selectivity we can obtain a clearly better result of the average relative error with the exception of query \emph{F3.3}. The result of this query can be approximated precisely by all approaches but the random sampling approaches fail to provide full bin completeness.

\begin{figure*}[!htb]
\centering
\begin{subfigure}{0.32\textwidth}
\centering
\includegraphics[width=\linewidth]{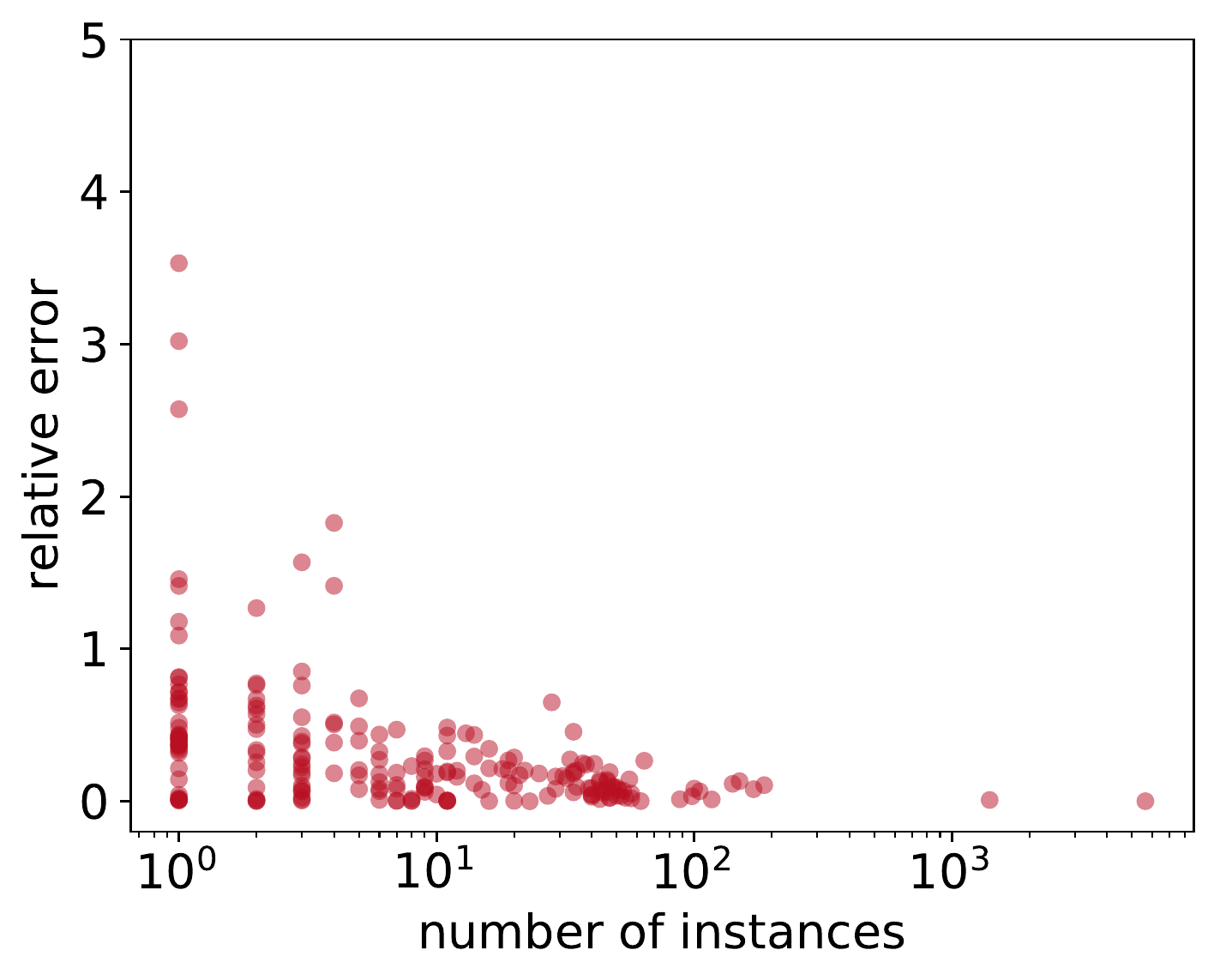}
\caption{SPN build on dataset \emph{flights100K}.}
\end{subfigure}
\begin{subfigure}{0.32\textwidth}
\centering
\includegraphics[width=\linewidth]{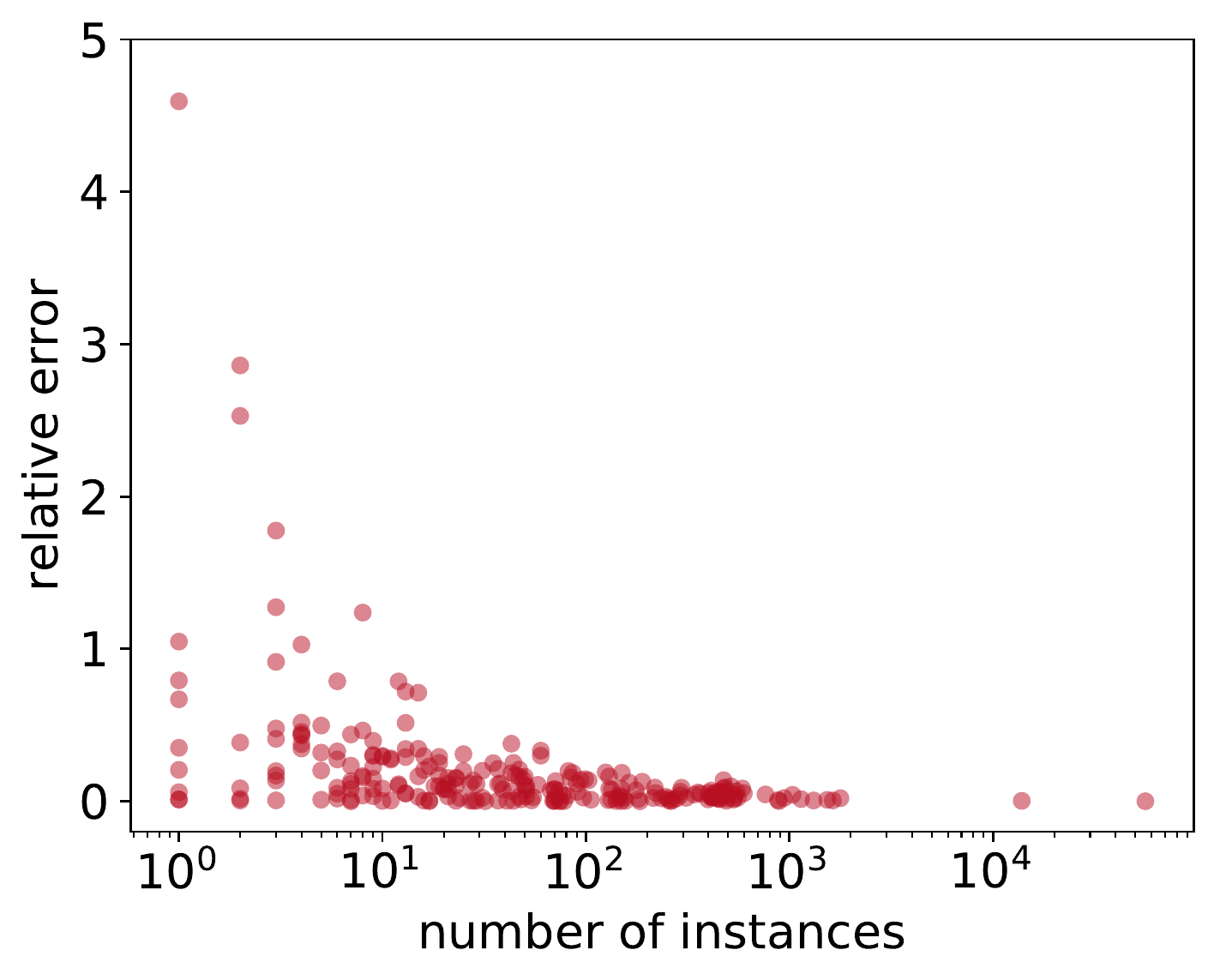}
\caption{SPN build on dataset \emph{flights1M}.}
\end{subfigure}
\begin{subfigure}{0.32\textwidth}
\centering
\includegraphics[width=\linewidth]{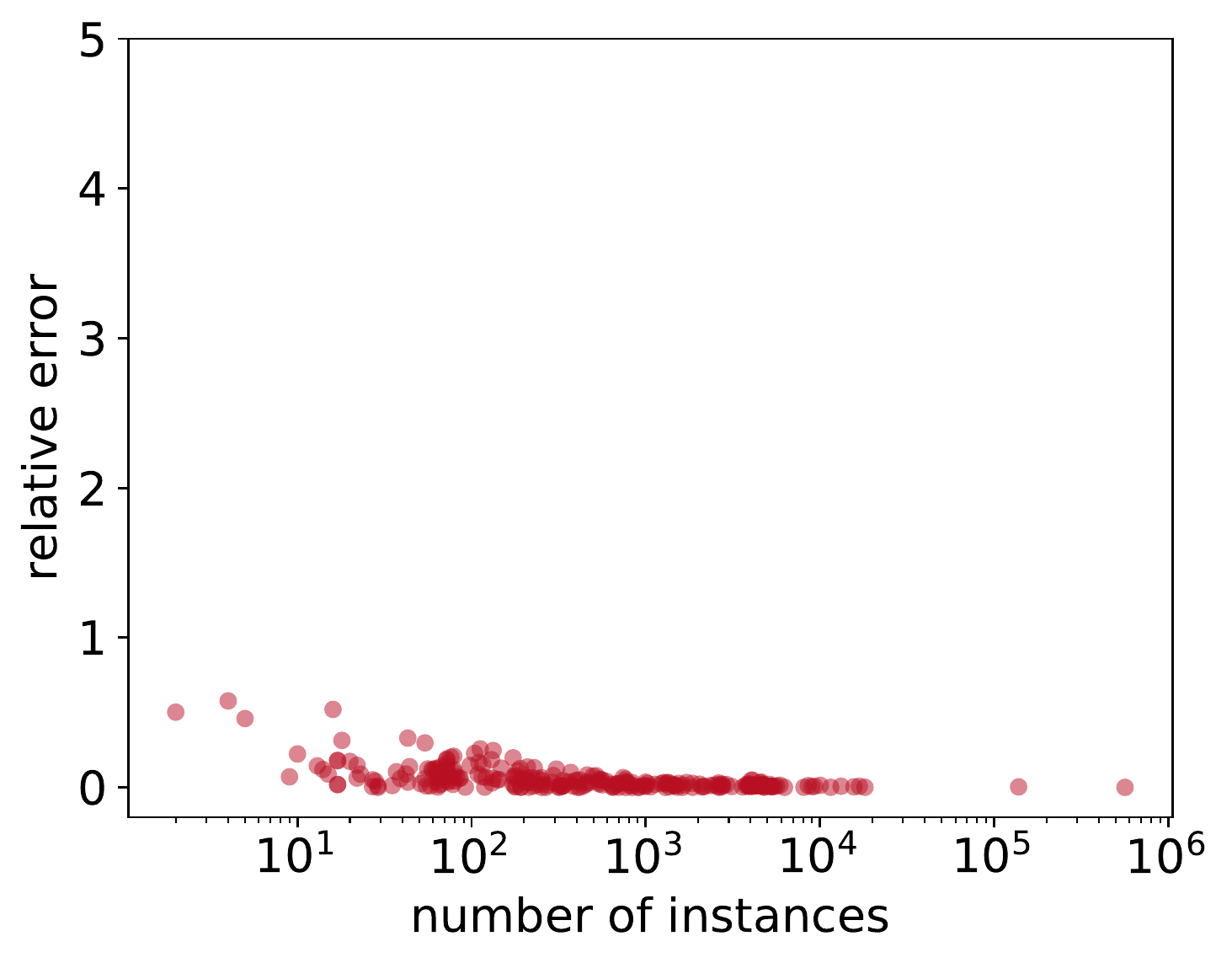}
\caption{SPN build on dataset \emph{flights10M}.}
\end{subfigure}
\caption{Relative error for all groups of all queries computed with the probability-based approach on different SPNs.}
\label{fig:scalability:rel_error_count}
\end{figure*}

\begin{figure*}[!htb]
\centering
\begin{subfigure}{0.245\textwidth}
\centering
\includegraphics[width=\linewidth]{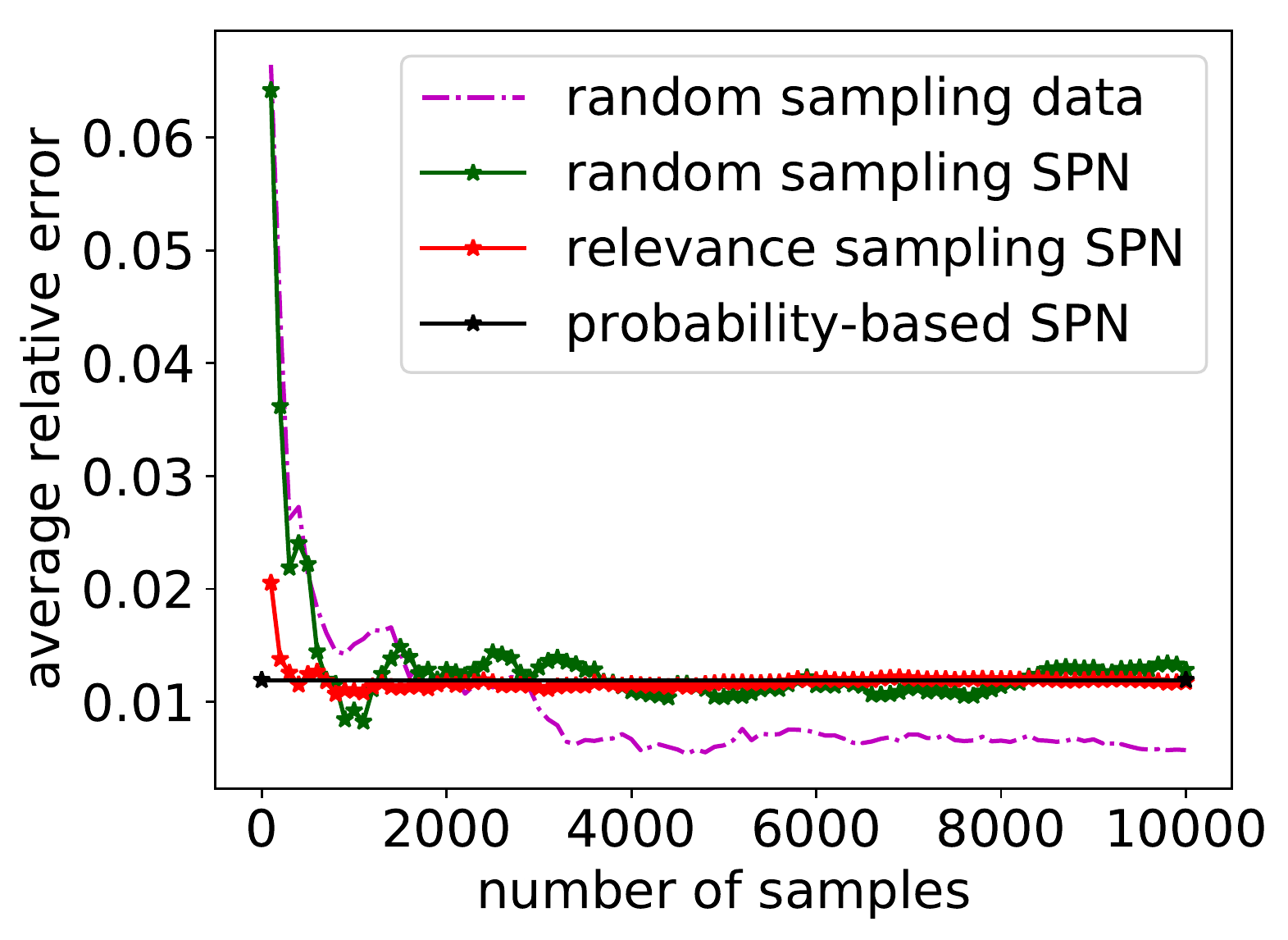}
\caption{S4.1 (selectivity $5\%$)}
\label{fig:evaluation:selectivity:results:1}
\end{subfigure}
\begin{subfigure}{0.245\textwidth}
\centering
\includegraphics[width=\linewidth]{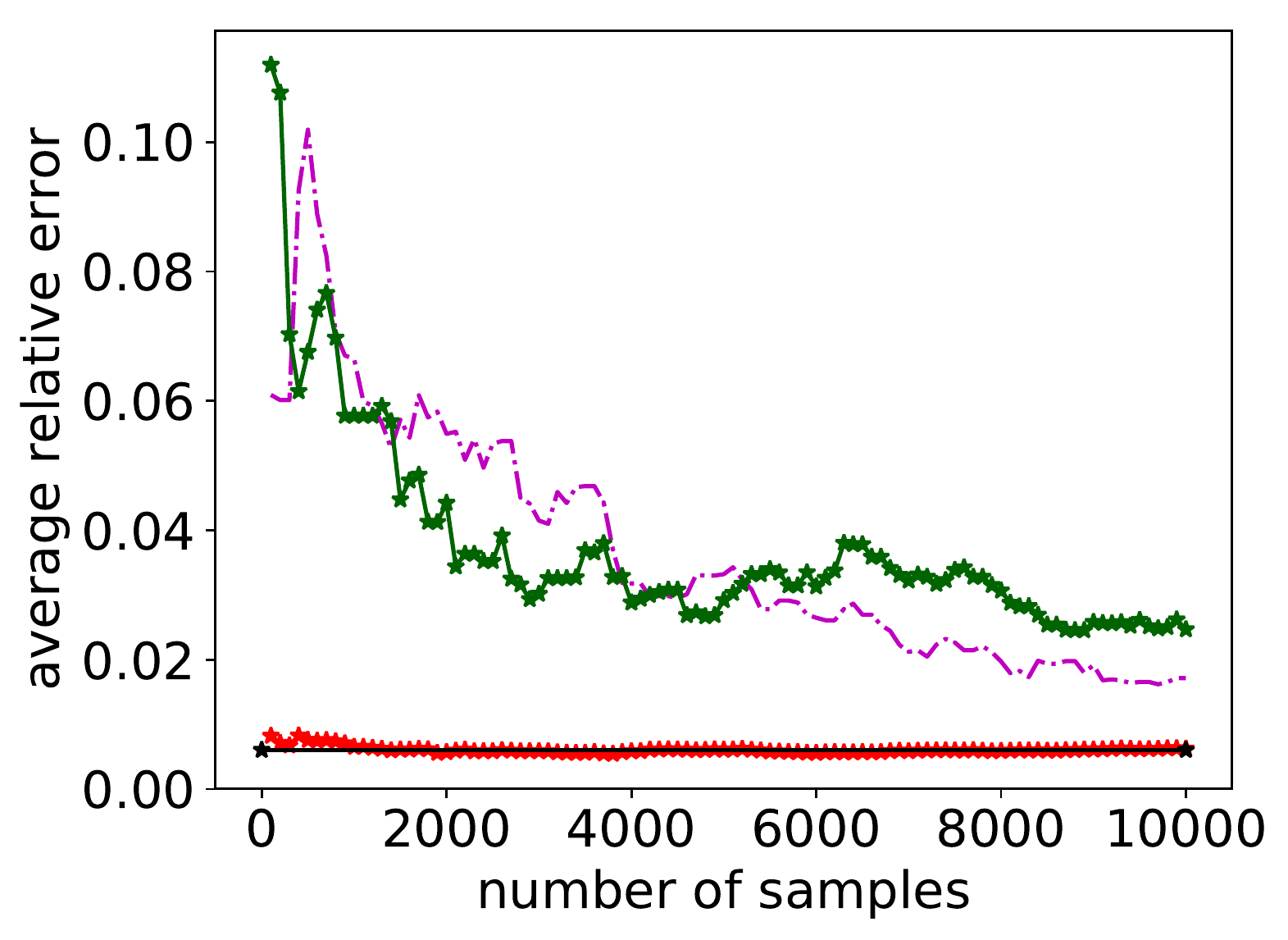}
\caption{S4.2 (selectivity $0.25\%$)}
\end{subfigure}
\begin{subfigure}{0.245\textwidth}
\centering
\includegraphics[width=\linewidth]{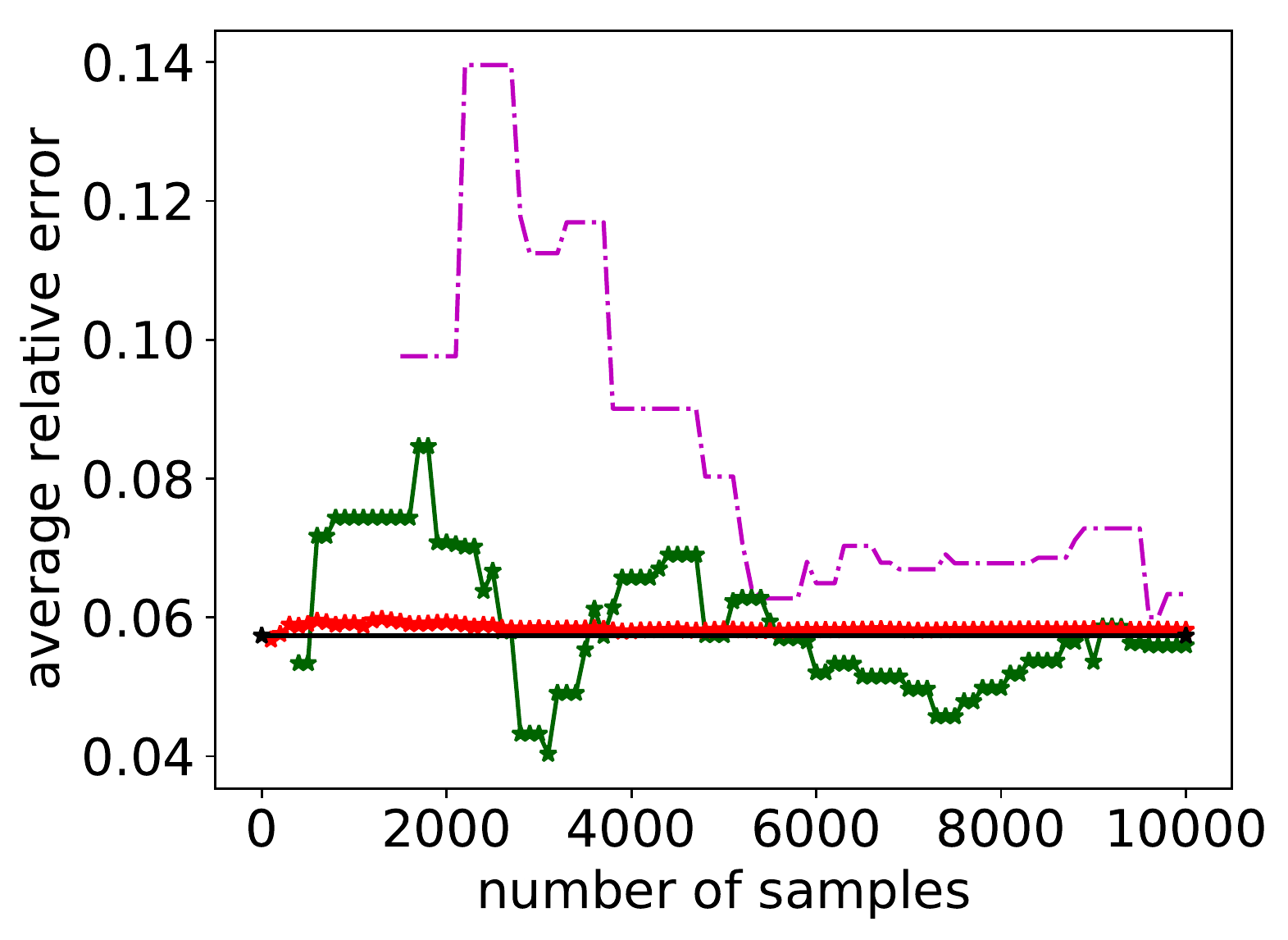}
\caption{S4.3 (selectivity $0.025\%$)}
\end{subfigure}
\begin{subfigure}{0.245\textwidth}
\centering
\includegraphics[width=\linewidth]{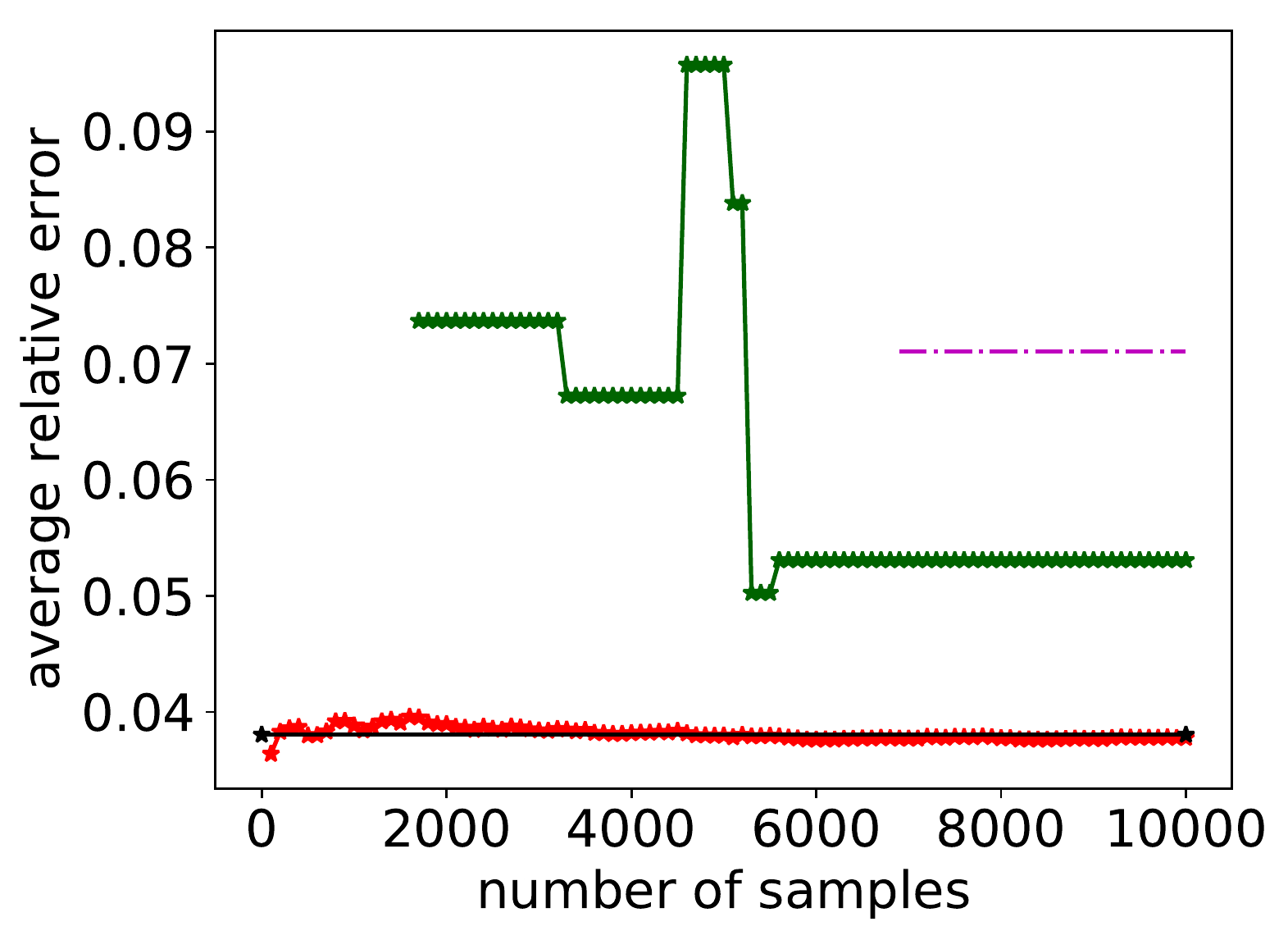}
\caption{S4.4 (selectivity $0.0025\%$)}
\end{subfigure}
\caption{Relative error of the sample-based approaches on the \emph{synthetic} dataset with respect to the number of generated samples.}
\label{fig:evaluation:selectivity:results}
\end{figure*}

\subsection{Exp. 2: Scalability}

Another important question regarding our proposed AQP techniques is how they behave when they are applied on different sizes of datasets. Therefore, we have used the dataset \emph{flights} with $100,000$, $1,000,000$ and $10,000,000$ instances on which we have run all ten queries. Furthermore, we have evaluated random sampling from data and exact SQL in order to draw comparisons. As for the previous experiment, we produce samples with the sample-based approaches until the average relative error is below $5$\% and bin completeness is fulfilled with the constraint that a maximum of $100,000$ samples can be generated. Regarding the exact SQL and the probability-based approach, we only report the average relative error and the runtime of these approaches. 

Statistics about the constructed SPNs on the different datasets are displayed in Table~\ref{tab:evaluation:scalability:spn_statistics}. Here, we can observe that we can obtain a very good compression of the data, especially for the bigger datasets. In particular, all SPNs share a similar amount of nodes which is caused by the same parameter configurations.

\begin{table}[!t]
\centering
\footnotesize
\caption{Statistics about the constructed SPNs for different sizes of the dataset \emph{flights}.}
\label{tab:evaluation:scalability:spn_statistics}
\begin{tabular}{rrrrrrrrr}
\toprule
  \multicolumn{1}{c}{\textbf{Dataset}}  &       \multicolumn{1}{c}{\textbf{Construction time }} &       \multicolumn{1}{c}{\textbf{Relative size of}}  &      \multicolumn{1}{c}{\textbf{Number of}}  \\
  &\multicolumn{1}{c}{\textbf{in minutes}}  &\multicolumn{1}{c}{\textbf{SPN to data}} &\multicolumn{1}{c}{\textbf{SPN nodes}}\\ 

\midrule
 flights100K &          $\sim$~~80 &       5.17\% &        901\\
 flights1M &         $\sim$~174 &         1.20\% &        802 \\
 flights10M &         $\sim$~248 &        0.32\% &        850 \\
\bottomrule
\end{tabular}
\end{table}

First of all, we examine the impact of the different dataset sizes on the execution time of the approaches. We have visualized the runtime for some selected queries in a grouped bar plot which can be seen in Figure~\ref{fig:scalability:bar_plot}. Each group represents a specific size of the dataset while the y-axis represents the execution time.

For exact SQL and random sampling from data, we can observe that the runtime for each different query on a particular dataset is similar whereas the execution time between the different datasets rises with respect to the size of the dataset linearly. 
This behavior can not be observed for the proposed AQP-techniques. In particular, we only detect a minor increase of the runtime on bigger datasets in the case that the query contains continuous columns (e.g. query \emph{F1.2} and \emph{F4.2}). This can be explained by the fact that the constructed SPNs for the different datasets contain almost the same amount of nodes. Therefore, the constructed SPNs mainly differ in the number of values which are stored in the leaves for continuous random variables. In the case of the bigger dataset, more values need to be accessed in the leaves to process the query. Thus the computation time is higher. In contrast, the execution time of queries on only discrete columns is not affected by the size of the dataset (e.g. query \emph{F2.3}).
As observed in the previous experiment, we can also observe that the execution time of the stratified sampling and the probability-based approach relies on the number of groups of the query (e.g. query \emph{F3.2}).

Apart from the execution time of the approaches, we can observe a major improvement in terms of the quality of the SPN with an increasing size of the dataset. Indeed, the average relative error for all evaluated queries decreases drastically with an increasing size of the dataset. Since the probability-based approach can be used to measure the quality of the SPN we have displayed the result of this approach for all queries and all datasets in Table~\ref{tab:scalability:error}.

In order to understand the reason for the improvement of the quality, we have analyzed the approximation for each group of the queries in more detail. We have computed the relative error for each particular group with the probability-based approach and set it into relation with the number of instances for that group in the actual dataset. We have done this procedure for all three datasets and visualized the results in Figure~\ref{fig:scalability:rel_error_count}. Here, the y-axis represents the relative error of the group and the x-axis represents number of instances of that group in the actual dataset.

In the plots, we can observe that the relative error decreases significantly if the group contains more instances. Furthermore, we can see that the groups which are computed on the bigger datasets contain much more instances than the groups of the smaller datasets. 

However, the relation between the relative error and the number of instances for the groups among the results on the different SPNs is the same. In particular, all groups which contain more than $100$ instances always have a relative error below $30$\%. Hence, we are able to obtain significantly better results with the SPN which is build on the dataset with ten million instances because the queried sub-population for a particular group contains more instances. In contrast, the groups of the dataset which rely on $100,000$ instances often contain only one or two instances for each group. Therefore, we can conclude the relative error for a group computed with the probability-based approach highly depends on the number of instances for that group. To sum it up, the quality of the approximations computed with an SPN is independent of the size of the dataset on which it is built and mainly depends on the number of instances of the groups of a query.

\begin{table*}[!t]
\centering
\caption{The average relative error of the probability-based approach for the queries \emph{S3} and \emph{S4} on the \emph{synthetic} dataset.}
\label{tab:evaluation:model_quality:syn_results}
\footnotesize
\begin{tabular}{llrrrrrrrrrr}
\toprule
  setting  &       \textbf{S 3.1} &       \textbf{S 3.2} &       \textbf{S 3.3} &      \textbf{ S 3.4} &       \textbf{S 4.1} &       \textbf{S 4.2} &       \textbf{S 4.3} &       \textbf{S 4.4} \\
\midrule
SPN used in previous experiments & 0.0000 & 0.1022 &  0.7404 &   2.9326 & 0.0119 & 0.0061 & 0.0574 & 0.0380 \\
SPN which assumes independence on all columns & 0.7367 & 4.3573 & 51.2983 & 597.7825 & 0.2391 & 0.2408 & 0.2376 & 0.2183 \\
\bottomrule
\end{tabular}
\end{table*}

\begin{figure*}[!t]
\begin{subfigure}{0.49\textwidth}
\centering
\includegraphics[width=0.49\linewidth]{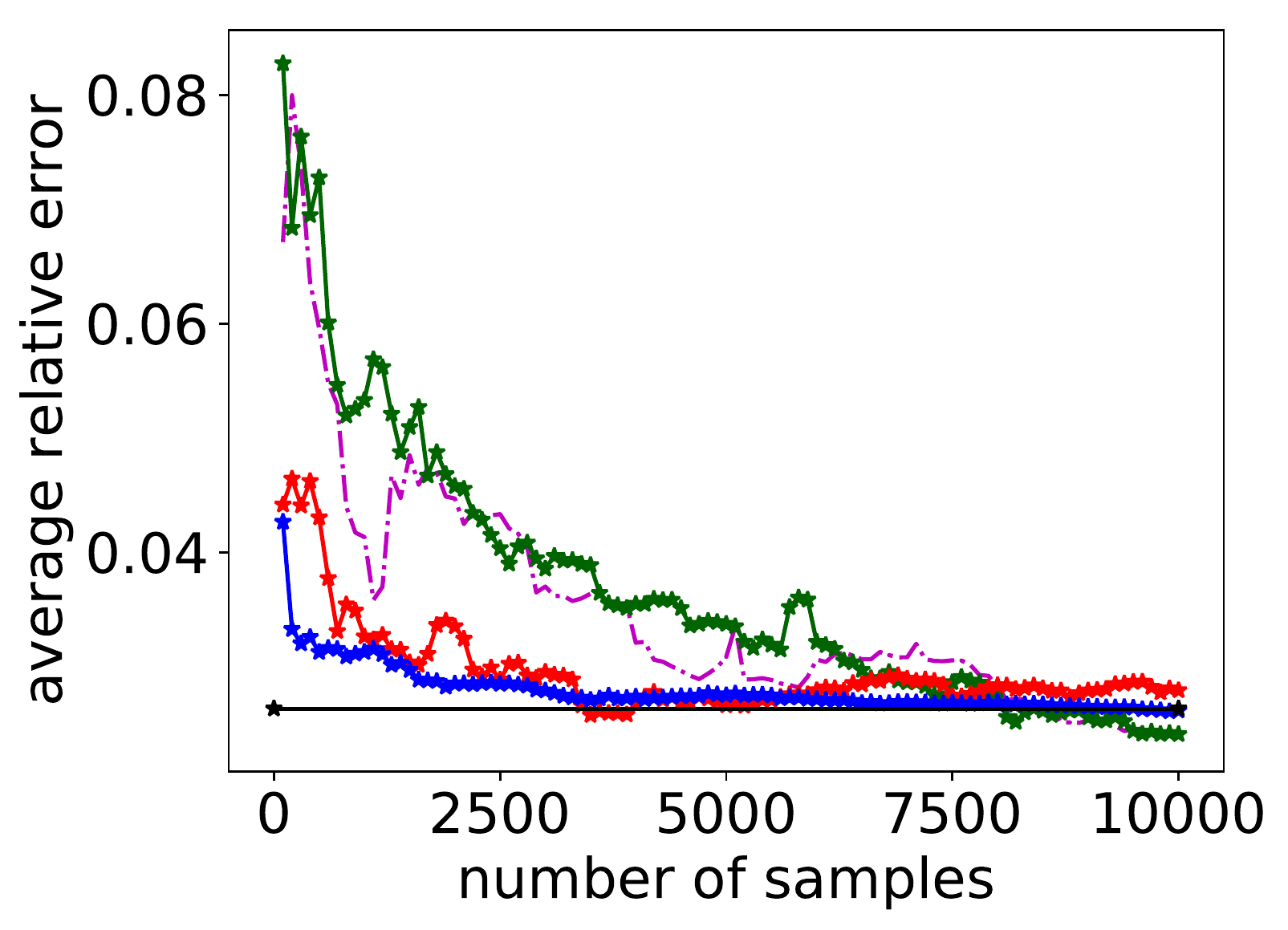}
\includegraphics[width=0.49\linewidth]{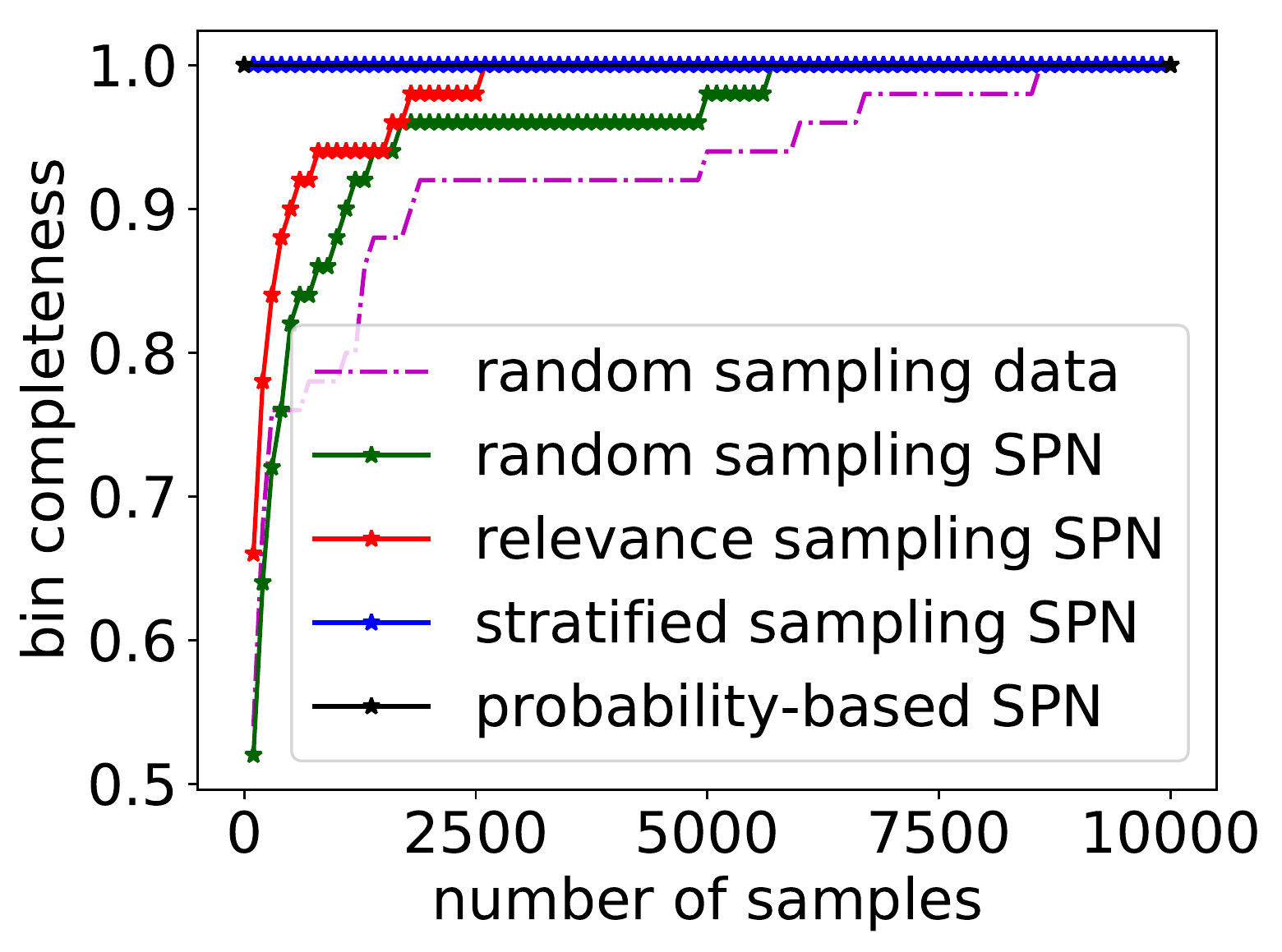}
\caption{Result for query \emph{S2.3}.}
\end{subfigure}
\begin{subfigure}{0.49\textwidth}
\centering
\includegraphics[width=0.49\linewidth]{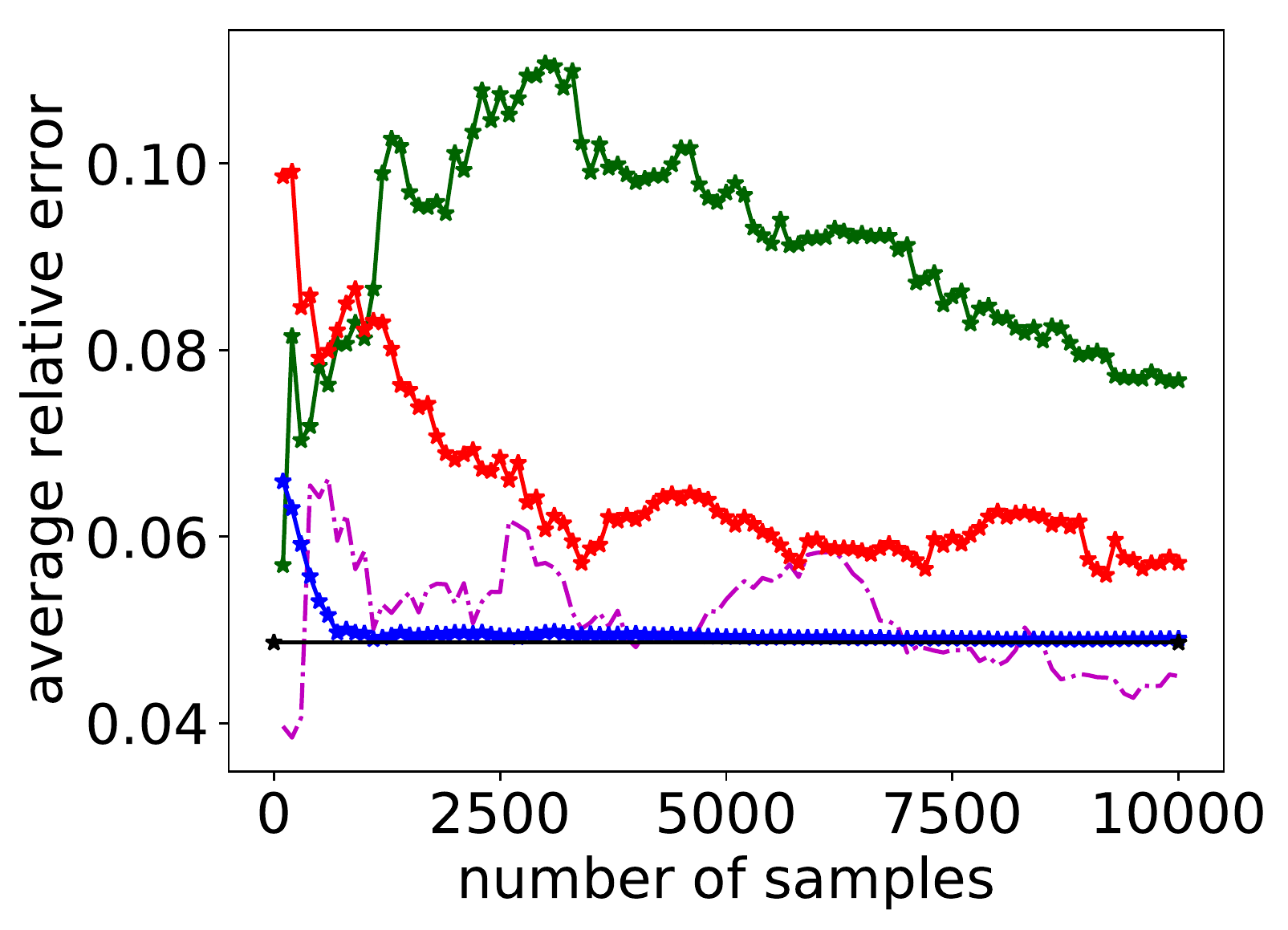}
\includegraphics[width=0.49\linewidth]{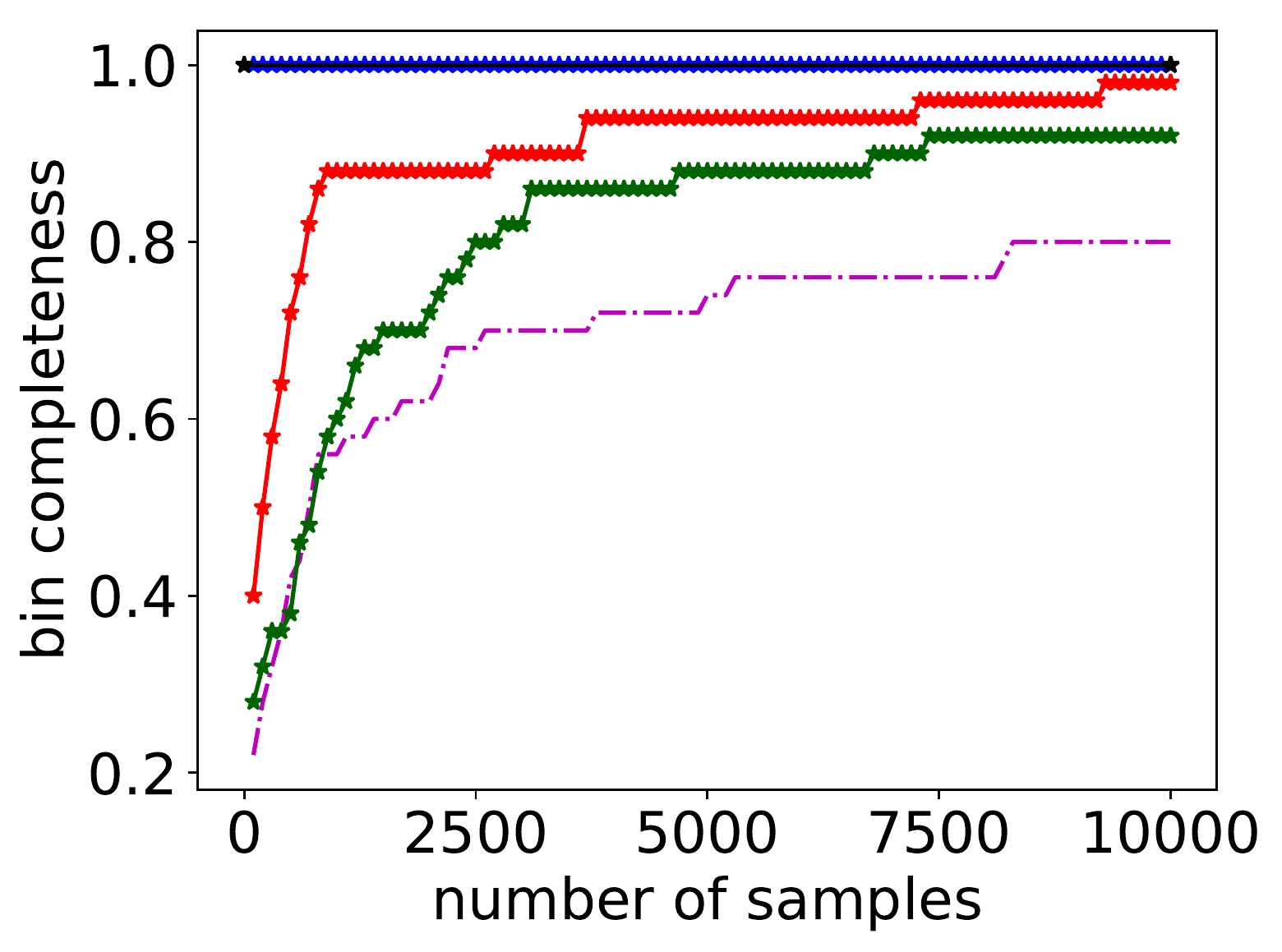}
\caption{Result for query \emph{S2.4}.}
\label{fig:evaluation:skewness:results:3}
\end{subfigure}
\caption{Results of the sample-based approaches on the \emph{synthetic} dataset with respect to the number of generated samples.}
\label{fig:evaluation:skewness:results}
\end{figure*}

\subsection{Exp. 3: Effect of Selectivity}

The advantage of our proposed biased sampling approaches with respect to the selectivity of a query is examined in this experiment. In order to directly measure the influence of the selectivity on the quality of the approximations we have evaluated the queries \emph{S3} and \emph{S4} with our proposed approaches on the \emph{synthetic} dataset. For direct comparison, we additionally evaluated random sampling from data. For all the sample-based approaches, we have generated $10,000$ samples and tracked the relative error of the approximation along the sample generation. Since the evaluated queries do not contain a grouping, the relevance and the stratified sampling approach are identical. Thus, we have not evaluated the stratified sampling approach for this experiment.

The first observation is that the result of the relevance sampling approach and the probability-based approach are the same for the query \emph{S3} irrespective of the number of generated samples. The reason for this behaviour is that the scaling of the \emph{COUNT} aggregation for the relevance sampling approach corresponds to the computation of the \emph{COUNT} aggregation with the probability-based approach.

The results for queries $S4.1$ to $S4.4$ are visualized in Figure~\ref{fig:evaluation:selectivity:results} where the x-axis represents the generated number of samples and the y-axis represents the relative error of the approximation. For clarity, we have visualized the result of the probability-based approach as a horizontal line in the plots. 

Compared to random sampling, we can observe in the plots that we are able to achieve good results with the relevance sampling approach after obtaining only a few samples. As aforementioned, the quality of the approximations of our proposed approaches mainly depend on the quality of the SPN. Therefore, the result of the relevance sampling approach always converges to the result of the probability-based approach and does not improve with more samples. This is clearly visible for the high selectivity query displayed in Figure~\ref{fig:evaluation:selectivity:results:1} in which random sampling outperforms the SPN approaches with increasing number of samples. However, the results of the low selectivity queries demonstrate the advantage of relevance sampling. In fact, the relevance sampling approach is independent of the selectivity of the query because it does not discard any samples, but the quality of the approximations is bounded to the quality of the SPN.

\subsection{Exp. 4: Effect of Skewness}

With the same setting as for the previous experiment, we examine the skewness of the grouping on our proposed approaches. Therefore, we now include the stratified sampling approach and evaluate the queries \emph{S1} and \emph{S2} on the \emph{synthetic} dataset.

The first and foremost observation on all queries is that the relevance and the stratified sampling approach are able to obtain better results with fewer samples compared to random sampling. The reason for this behaviour is that the biased sampling approaches are able to use all the generated samples while the random sampling approaches have to discard approximately $75\%$ of the generated samples due to the selectivity of the queries.

Figure~\ref{fig:evaluation:skewness:results} visualizes our results examplarily for the queries \emph{S2.3} and \emph{S2.4}. The x-axis represents the generated number of samples and the y-axis represents the value for average relative error or the bin completeness. The result of the probability-based approach is visualized as a horizontal line in the plots. 

In the plots, we can observe that the stratified sampling approach instantaneously reaches full bin completeness irrespective of the skewness of the SQL-query. In contrast, the random sampling approaches and even the relevance sampling approach require a considerable amount of samples to provide approximations for all the groups. The fact that random sampling obtains a respectable good average relative error on query \emph{S2.4} is caused by the low bin completeness since the missing approximations for the low selectivity are not considered by this measure.

\subsection{Exp. 5: Effect of Model Quality}

As already stated in the previous experiments, the quality of the approximations of our proposed AQP approaches depend on the quality of the SPN. The quality of an SPN is mainly defined by the parameters for the construction. Hence, we examine the effect of the parameter configuration for the construction of the SPN on the quality of the approximations in this experiment.  We have chosen to use the \emph{synthetic} and the \emph{flights1M} dataset on which we evaluated all queries with the probability-based approach. For the hyper-parameter configuration of the SPN, we varied the \emph{rdc-threshold} as well as the \emph{min instance slice} (for explanation of the parameters we refer to \cite{spn_mspn}).

The first observation on the results of the dataset \emph{flights1M} is that the different hyper-parameter settings for the construction of the SPN has only minor influence on the quality of the approximations. Regarding the most general SPN, which assumes independence on all columns, the quality of the SPN does not improve with more precise parameter settings. In contrast, on the synthetic dataset the parameter settings have a major influence on the quality of the computed approximations. The improvement of average relative error for queries \emph{S3} and \emph{S4} of the \emph{synthetic} dataset are displayed in Table~\ref{tab:evaluation:model_quality:syn_results}.

A closer examination of the results reveals that the relative error among the queries increases for a smaller sub-population on which the query is applied. In particular, for the queries \emph{S3.1} to \emph{S3.4} this observation is clearly visible.
As already investigated in the first experiment, the reason is that the computation for groups with only a few instances is more error prone than for groups with more instances. This explains the fact that queries on very rare sub-populations result in a higher average relative error. However, we are interested in why these approximations are likely to yield a higher relative error.

\begin{figure}[!t]
\centering
\begin{subfigure}{0.23\textwidth}
\centering
\includegraphics[width=\linewidth]{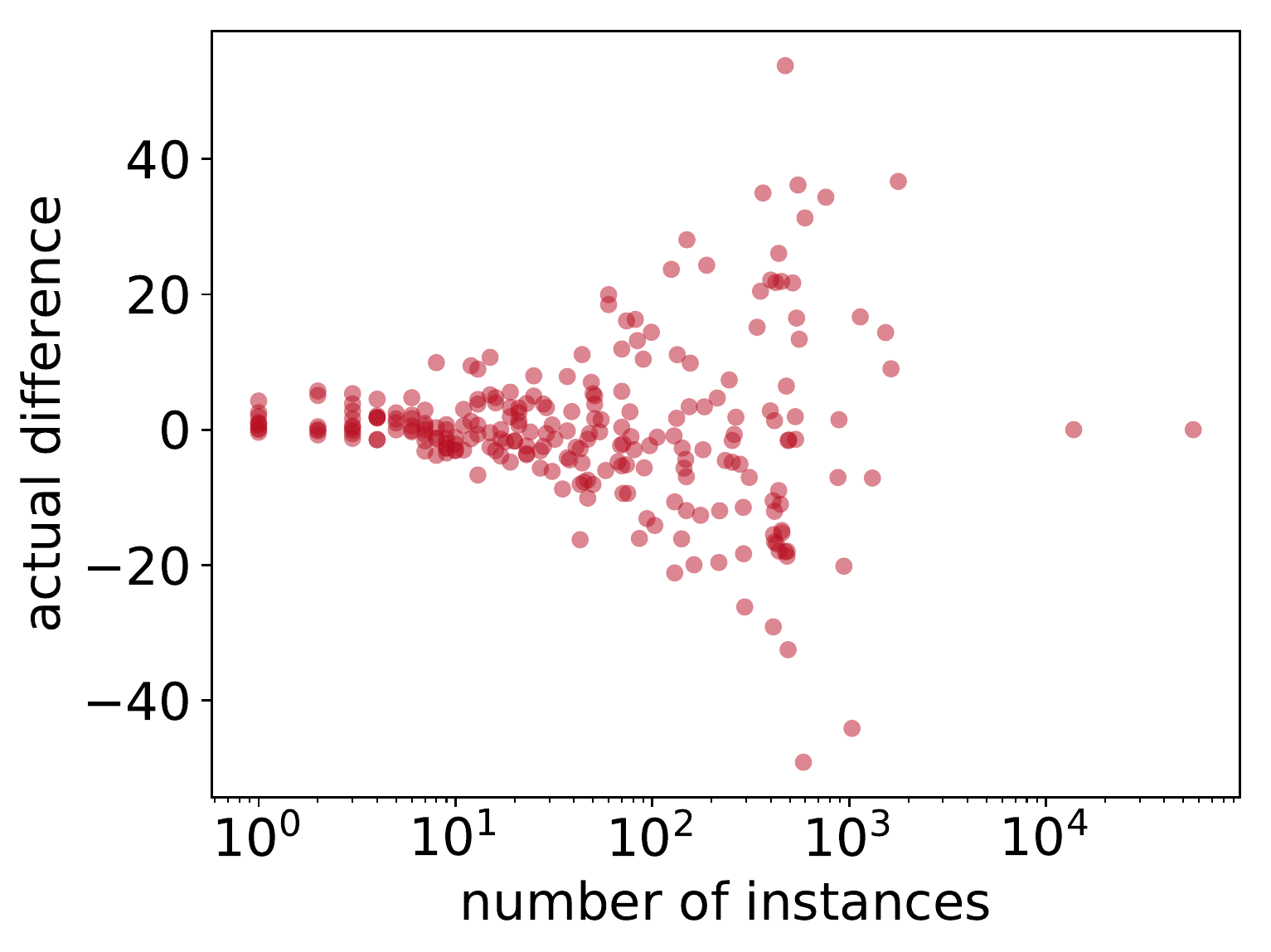}
\caption{Absolute difference.}
\label{why}
\end{subfigure}
\begin{subfigure}{0.23\textwidth}
\centering
\includegraphics[width=\linewidth]{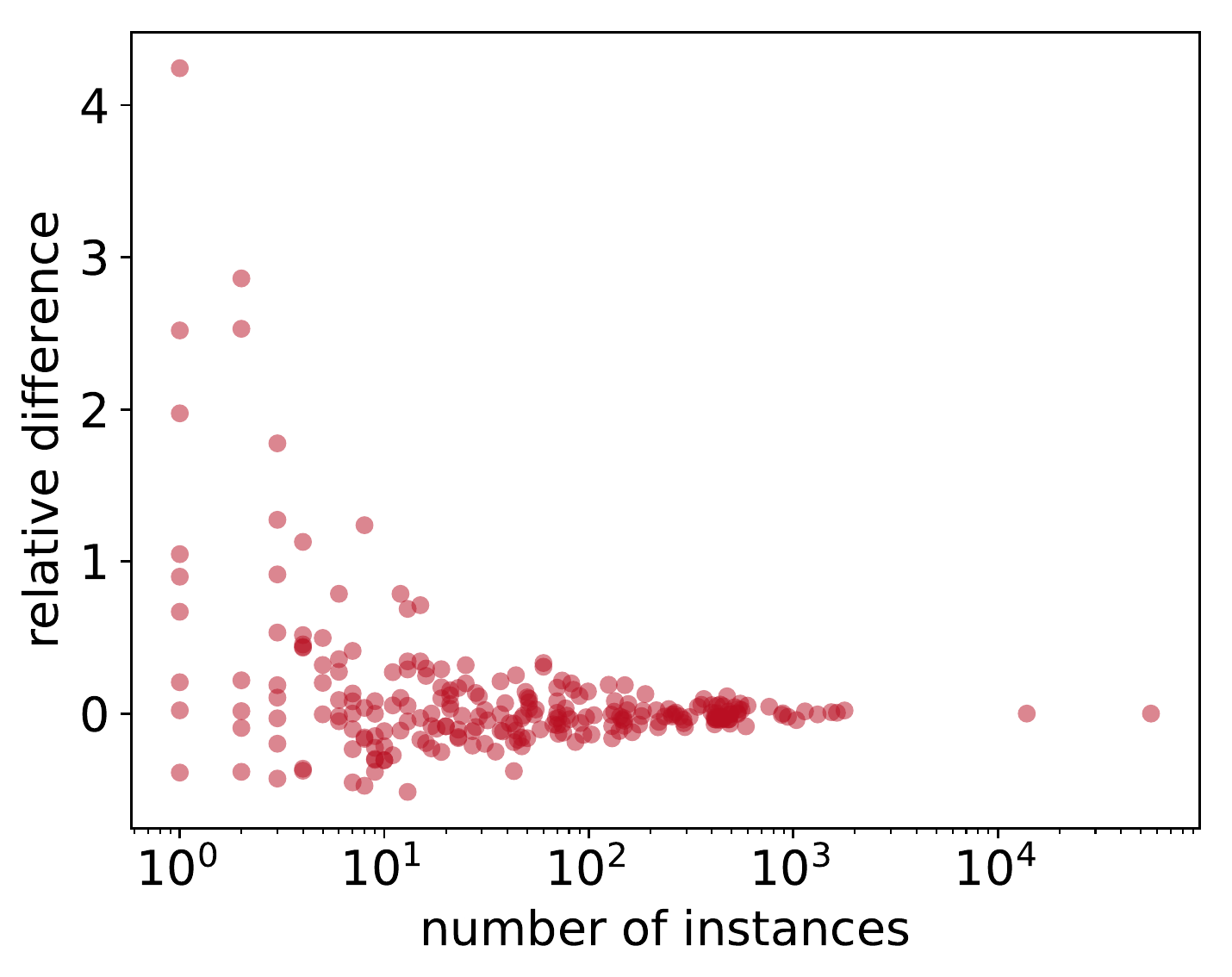}
\caption{Relative difference.}
\label{fig:evaluation:model_quality:rel_count_groups}
\end{subfigure}
\caption{Difference between the actual amount of instances in the dataset \emph{flights1M} and the computed amount of instances with SPN for all groups of all queries.}
\vspace{-3ex}
\label{fig:evaluation:model_quality:count_groups}
\end{figure}

Taking a closer look at the results of all queries, we recognize that this behavior only applies for the queries with \emph{COUNT} and \emph{SUM} aggregations. Since these aggregations need to be scaled up in order to obtain the approximation, they rely on the computation of inference for the query which may be the cause for the error. 
Hence, we have analyzed the result of the inference computation in more detail on the dataset \emph{flights1M}. In particular, we have used the SPN to estimate the number of instances of the sub-population for each group of the queries and set it into relation with the actual number of instances for the respective group in the actual dataset. On the one hand, we have visualized the absolute difference which can be seen in Figure~\ref{why} and, on the other hand, we have visualized the relative difference in Figure~\ref{fig:evaluation:model_quality:rel_count_groups}. For both plots the x-axis represents the number of instances for the group in the actual dataset whereas the y-axis represents the (relative) difference of the estimated number of instances. A positive value for the y-axis implies that the SPN has overestimated the number of instances for a particular group while a negative values implies that the number of instances has been underestimated.

In Figure~\ref{why} we can observe that the computed number of instances with the SPN is actually very accurate for groups which contain only a few instances. In contrast, the groups containing more instances have a bigger error in the estimated number of instances. However, since we are dealing with the relative error, we have to examine the relative difference which is displayed in Figure~\ref{fig:evaluation:model_quality:rel_count_groups}. Here, we can observe that the relative difference is much higher for the groups which rely on only a few instances. Furthermore, the extent of the relative difference decreases with increasing size of the group even though the actual difference of the estimation gets bigger. Therefore, we can conclude that the high relative error is mainly caused by the high relative difference in the approximation for the number of instances for very small sub-populations. Moreover, this observation also explains the good results on query \emph{F3.3} and queries \emph{S4.1} to \emph{S4.4} because for the approximation of an \emph{AVG} aggregation, we do not have to compute the size of the sub-population with the SPN and, therefore, we do not have this bias in our approximation. 

This suggests that we can use the SPN to get a certainty about the approximated result. In particular, the SPN can be used to evaluate the quality for the approximation of a SQL query without accessing the actual data. We have to obtain the number of instances for the groups. In case that only a few instances are available for a specific group, we can conclude that the result may not be very precise. In contrast, we can be more certain about the approximation if the group contains more instances.

With the increase of the size of the SPN, we can observe that the runtime for the probability-based approach increases as well. Taking the scalability experiment into account, we can conclude that the execution time highly depends on the size of the SPN while the size of the dataset has only minor impact. 
Apart from the size of the SPN, we can see that the queries which rely on many columns have a significantly higher runtime than queries which are only applied on a few columns. The reason for this behavior is that the SPN is marginalized beforehand to the set of relevant columns. Due to the marginalization inference and expectation is computed a smaller SPN which saves a lot of computation time.
Since the runtime of the proposed AQP techniques matters, we have to choose a parameter configuration for the construction of the SPN which ensures that the size of the SPN does not grow to big and that the data is represented precisely. Our evaluation of different parameter settings has shown that the SPN can grow very big with a very precise parameter settings. However, this is not ideal because the runtime of the proposed AQP techniques increases rapidly. Thus, we have to find a trade-off between the accurateness and the size of the SPN.

%% file: sections/related_work.tex
\section{Related Work}
\label{sec:related}

Approximate query processing (AQP) emerged from the need to reduce the response time of queries executed on huge amounts of data. AQP gained more and more interest due to the technological advancements and cheap storage cost of data. Especially in recent years a lot of research took place in this field~\citep{aqp_liu}. In general, the primary focus of AQP techniques is the approximation of aggregation queries whereby \emph{COUNT}, \emph{AVG} and \emph{SUM} aggregations are the most popular. Ideally, the AQP approach can support all kinds of aggregation queries~\citep{aqp_database_learning, aqp_reuse} including joins and nested queries. Moreover, no assumptions about the data should be made in advance and the approximation of a query result should be a magnitude faster than executing the query on the whole table~\citep{aqp_database_learning}.

AQP techniques belong to one of the following two categories: (1) Sample-based approaches and (2) approaches which rely on a pre-computed synopsis. For sample-based approaches, we can further differentiate between unbiased (e.g. random sampling) and biased sampling (e.g., stratified sampling). Biased sampling approaches like \emph{BlinkDB}~\citep{aqp_blinkdb} or dynamic sample selection~\citep{aqp_dynamic_sample} can provide fast approximations over rare sub-populations in an efficient manner but need to maintain a set of selected samples in memory. In general, such sets of samples cannot be computed in advance since no prior knowledge about the workload is given~\citep{aqp_reuse, aqp_blinkdb}.
This problem is solved by obtaining the samples during query time~\citep{aqp_online_aggregation_base}. Most of the sample-based approaches, in particular the unbiased sampling techniques, are applied during query time on the fly which is referred as online aggregation. Systems for online aggregation which rely on unbiased samples, like CONTROL \citep{aqp_control} or DBO \citep{aqp_dobra}, typically provide good approximations for queries over the mass of the distribution but cannot provide fast answers for queries on rare sub-populations~\citep{aqp_reuse}.
The original work for online aggregation considered only queries on one table while follow-up work extended the techniques also for queries with joins \cite{aqp_wander_join,aqp_ripple_join}.

The difference between these approaches and our model-based AQP is as follows: Generally, in the classical AQP approaches prior knowledge about the workload is required in advance, to generate a set of biased samples. However, especially for interactive data exploration the queries are not known in advance, for which reason these samples are obtained during query time~\citep{aqp_reuse, aqp_blinkdb}. 
Different from those approaches, model-based AQP  can produce biased samples on-the-fly and thus supports ad-hoc queries over rare sub-populations.

Moreover, there are AQP approaches which construct a synopsis of the data in advance \cite{DBLP:conf/vldb/GarofalakisG01}. For example, such data structures can be materialized views or data cubes~\citep{aqp_data_cube}. Nevertheless, any other data structure can be used for the synopsis if can provide answers for aggregation queries. In general, approaches which rely on synopsis usually lack the ability to support all possible SQL-queries since they have abstracted the entries of the table. The most famous online analytic processing tool is \emph{OLAP}~\citep{aqp_olap_overview} for which a data cube is constructed by defining hierarchies over the dimensions. Using this data cube, aggregation queries over these hierarchical dimensions can be answered efficiently. However, OLAP provides poor performance for queries on rare sub-populations~\citep{aqp_blinkdb} and the computation of the cube can take significant processing time~\citep{aqp_online_aggregation_base}.

%% file: sections/conclusion.tex
\section{Conclusion and Future Work}
\label{sec:concl}

In this work, we have proposed a new approach for AQP using generative models. 
With our proposed approach, we are able to overcome major problems of classical AQP approaches.
In our experimental evaluation, we have shown that the accuracy of the approximations of our model-based approaches outperform classical AQP approaches on real and synthetic data sets. 
Furthermore, the runtime of our proposed approaches mainly depends on the size of the constructed SPN and not on the data set size making them an ideal candidate for exploring large data sets. 

For future work, it remains to analyze the behavior of the proposed approaches on other datasets. In particular, the model quality of the constructed SPNs needs to be examined in more detail. In addition, the quality of the SPNs for AQP could be further improved. In future we want to introduce SPNs that are a hybrid of the data and the model and keep the original data if high correlations between attributes are detected. By doing this, we can avoid losing particular correlations among the data for sub-populations which are difficult to separate. More precisely, we avoid creating product nodes which introduce a major error into the SPN. 
Furthermore, we have proposed a way to compute confidence bounds with the SPN and a way how disjunctive conditions among different columns can be handled. 
Finally, in future we plan to also support more complex queries including joins and nested queries. The general idea is to construct a SPN for each table individually and to combine the approximations of the different SPNs during query runtime.

%% file: sections/appendix.tex
\section*{Appendix}
\label{sec:appendix}

\begin{lstlisting}[language=SQL,escapechar=@,language=SQL,basicstyle=\scriptsize, caption={Queries for the dataset flights.}, label={dsdsdddd2}]
F 1.1 SELECT AVG(dep_delay) FROM flights WHERE origin='ATL'
F 1.2 SELECT AVG(distance) FROM flights WHERE unique_carrier='TW'
F 2.1 SELECT unique_carrier, COUNT(*) FROM flights
      WHERE origin_state_abr='LA' GROUP BY unique_carrier
F 2.2 SELECT unique_carrier, COUNT(*) FROM flights
      WHERE origin_state_abr='LA' AND  dest_state_abr='CA'
      GROUP BY unique_carrier
F 2.3 SELECT year_date, COUNT(*) FROM flights
      WHERE origin_state_abr='LA' AND dest='JFK' 
      GROUP BY year_date
F 3.1 SELECT year_date, SUM(distance) FROM flights
      WHERE unique_carrier='9E' 
      GROUP BY year_date
F 3.2 SELECT origin_state_abr, SUM(air_time) FROM flights
      WHERE dest='HPN' GROUP BY origin_state_abr
F 3.3 SELECT unique_carrier, AVG(dep_delay) FROM flights
      WHERE year_date='2005' AND origin='PHX'
      GROUP BY unique_carrier
F 4.1 SELECT dest_state_abr, COUNT(*) FROM flights
      WHERE distance>2500 GROUP BY dest_state_abr
F 4.2 SELECT unique_carrier, COUNT(*) FROM flights
      WHERE air_time>1000 AND dep_delay>1500
      GROUP BY unique_carrier	
\end{lstlisting}

\begin{lstlisting}[language=SQL,escapechar=@,language=SQL,basicstyle=\scriptsize, caption={Queries for the synthetic dataset.}, label={dsdsdddd2}]
S 1.1 SELECT A, COUNT(*) FROM syn WHERE filter='1' GROUP BY A
S 1.2 SELECT A, COUNT(*) FROM syn WHERE filter='2' GROUP BY A
S 1.3 SELECT A, COUNT(*) FROM syn WHERE filter='3' GROUP BY A
S 1.4 SELECT A, COUNT(*) FROM syn WHERE filter='4' GROUP BY A
S 2.1 SELECT A, AVG(B)   FROM syn WHERE filter='1' GROUP BY A
S 2.2 SELECT A, AVG(B)   FROM syn WHERE filter='2' GROUP BY A
S 2.3 SELECT A, AVG(B)   FROM syn WHERE filter='3' GROUP BY A
S 2.4 SELECT A, AVG(B)   FROM syn WHERE filter='4' GROUP BY A
S 3.1 SELECT    COUNT(*) FROM syn WHERE filter='1' AND A='4'
S 3.2 SELECT    COUNT(*) FROM syn WHERE filter='2' AND A='4'
S 3.3 SELECT    COUNT(*) FROM syn WHERE filter='3' AND A='4'
S 3.4 SELECT    COUNT(*) FROM syn WHERE filter='4' AND A='4'
S 4.1 SELECT    AVG(B)   FROM syn WHERE filter='1' AND A='4'
S 4.2 SELECT    AVG(B)   FROM syn WHERE filter='2' AND A='4'
S 4.3 SELECT    AVG(B)   FROM syn WHERE filter='3' AND A='4'
S 4.4 SELECT    AVG(B)   FROM syn WHERE filter='4' AND A='4'
\end{lstlisting}